\documentclass{mnras}

\usepackage{newtxtext,newtxmath}
\usepackage[T1]{fontenc}
\usepackage{comment}

\DeclareRobustCommand{\VAN}[3]{#2}
\let\VANthebibliography\thebibliography
\def\thebibliography{\DeclareRobustCommand{\VAN}[3]{##3}\VANthebibliography}

\usepackage{graphicx}	
\graphicspath{{./figures/}}
\usepackage{float}
\usepackage[dvipsnames]{xcolor}
\let\bibhang\relax
\usepackage{natbib}
\usepackage{subfig}
\usepackage{gensymb}
\usepackage{amsmath}
\usepackage{orcidlink}

\title[Spectroscopic Variations in J091016.43+210554.2]{Discovery of a Magnetic Double-Faced DBA White Dwarf}

\author[Moss et al.]{
Adam Moss$^{1}$\orcidlink{0000-0001-7143-0890},
P. Bergeron$^{2}$\orcidlink{0000-0003-2368-345X},
Mukremin Kilic$^{1}$\orcidlink{0000-0001-6098-2235},
Gracyn Jewett$^{1}$,
Warren R. Brown$^{3}$\orcidlink{0000-0002-4462-2341},
\newauthor{Alekzander Kosakowski$^{4}$\orcidlink{0000-0002-9878-1647},
and Olivier Vincent$^{2}$\orcidlink{0000-0002-7729-484X}}
\\
$^{1}$Homer L. Dodge Department of Physics and Astronomy, University of Oklahoma, 440 W. Brooks St., Norman, OK 73019, USA\\
$^{2}$Département de Physique, Université de Montréal, C.P. 6128, Succ. Centre-Ville, Montréal, Québec H3C 3J7, Canada\\
$^{3}$Center for Astrophysics | Harvard \& Smithsonian, 60 Garden Street, Cambridge, MA 02138, USA\\
$^{4}$Department of Physics and Astronomy, Texas Tech University, 2500 Broadway Lubbock, Texas 79409, USA\\
}

\date{Accepted 2023 December 9}
\pubyear{2023}

\begin{document}
\label{firstpage}
\pagerange{\pageref{firstpage}--\pageref{lastpage}}
\maketitle

\begin{abstract}

We report the discovery of spectroscopic variations in the magnetic DBA white dwarf SDSS J091016.43+210554.2. Follow-up time-resolved spectroscopy at the Apache Point Observatory (APO) and the MMT show significant variations in the H absorption lines over a rotation period of 7.7 or 11.3 h. Unlike recent targets that show similar discrepancies in their H and He line profiles, such as GD 323 and Janus (ZTF J203349.8+322901.1), SDSS J091016.43+210554.2 is confirmed to be magnetic, with a field strength derived from Zeeman-split H and He lines of $B\approx0.5$ MG. Model fits using a H and He atmosphere with a constant abundance ratio across the surface fail to match our time-resolved spectra. On the other hand, we obtain excellent fits using magnetic atmosphere models with varying H/He surface abundance ratios. We use the oblique rotator model to fit the system geometry. The observed spectroscopic variations can be explained by a magnetic inhomogeneous atmosphere where the magnetic axis is offset from the rotation axis by $\beta = 52\degree$, and the inclination angle between the line of sight and the rotation axis is $i = 13$ - $16\degree$. This magnetic white dwarf offers a unique opportunity to study the effect of the magnetic field on surface abundances. We propose a model where H is brought to the surface from the deep interior more efficiently along the magnetic field lines, thus producing H polar caps.
\end{abstract}
\begin{keywords}
stars: evolution — stars: rotation — white dwarfs — stars: magnetic fields — starspots
\end{keywords}

\section{Introduction}
While H- and He-dominated atmosphere white dwarfs (classified as DAs and DBs if they display neutral H and He lines, respectively) make up the majority of white dwarfs, their distribution across effective temperatures varies. \citet{Green86} found an extreme lack of DBs between $T_{\rm eff}\sim45,000$ K and 30,000 K, in the so-called ``DB gap''. While \citet{Eisenstein06} found multiple hot DBs in this gap from the Sloan Digital Sky Survey (SDSS), there was still a lack of DBs in this range by a factor of 2.5 compared to the number at 20,000 K. This suggests that there are various mechanisms that alter the atmospheric compositions of white dwarfs as they evolve along the cooling sequence.

The currently accepted model to solve the DB gap problem is the float-up model, originally proposed by \citet{Fontaine87}.  In this model, diluted H in the envelope diffuses upward, converting a DO or DB white dwarf into a DA star by the time the star cools to $T_{\rm eff}\sim45,000$ K. After this object reaches $\sim$30,000 K, convective dilution would cause the underlying He convection zone to erode the superficial  H outer layer, transforming the object into a DB or DBA white dwarf (see \citealt{Bedard22} for a more detailed description of this float-up model, and the spectral evolution of white dwarfs in general).

At lower temperatures, it is expected that this convective dilution process gives rise to the cool ($20,000\ {\rm K}\gtrsim T_{\rm eff} \gtrsim 12,000$ K) He-dominated, H-bearing white dwarfs (DBAs). These mixed atmospheres make up a large portion of the DB class \citep{Koester15,Rolland18}, but modelling the dilution process that generates these DBAs often leads to theoretical H abundances orders of magnitude lower than observed \citep{Macdonald91,Rolland18, Rolland20}. To solve this issue, \citet{Rolland20} invoked a dredge-up process in which the mixed H/He convection zone sinks deep into the star as dilution begins, resulting in a significant dredge-up of H from a deep reservoir. \citet{Bedard23} successfully reproduced the range of H abundances in the atmospheres of cool DBAs using only this internal reservoir of H as opposed to an external source such as accretion of planetesimals \citep{Farihi13,Gentile17}. 

From an observational standpoint, the presence of both H and He in a spectrum should be a clear indicator of a DBA (or DAB depending on which lines are stronger). However, for some targets it is difficult to adequately reproduce the H/He lines using a single-star, homogeneous (constant H/He abundance ratio across the surface) atmosphere model. \cite{Genest19b} analyzed 1915 DB white dwarfs from the SDSS and identified 10 unresolved DA+DB binary candidates based on their poor fits using a single-star model. This binary possibility can be invoked for other spectral types as well. \citet{Rolland15} analyzed 16 cool magnetic DA white dwarfs and found that 10 have weaker than expected H lines. To solve this, \citet{Rolland15} proposed that these 10 are in an unresolved DA+DC binary, effectively diluting the H$\alpha$ line profile.

\citet[][also see \citealt{Kilic19}]{Moss23} obtained follow-up time-series spectroscopy on eight of the targets from \citet{Rolland15} and found that five require a patchy surface composition to explain the shallow H$\alpha$ lines. There are no signs of significant radial velocity shifts in the time-series data either. It is likely that the magnetic field inhibits convection in regions where the field is stronger, creating a patchy atmosphere where the H abundance is higher or lower in different regions. \citet{Tremblay15} showed that a field strength of only $\sim$50 kG is needed to suppress convection, which is much lower than the $\sim$MG fields found in the \cite{Moss23} sample. 

Given this finding, it is likely that there are other objects with inhomogeneous atmospheres posing as unresolved double degenerates, but of different spectral types. These atmospheres would have varying H/He abundance ratios across the stellar surface, which would manifest in time-series spectroscopy as varying absorption lines. In this paper we report the discovery of a magnetic DBA white dwarf with an inhomogeneous surface abundance, which was initially classified as an unresolved DA+DB binary by \cite{Genest19b}. We first present in Section 2 evidence based on published spectroscopic and photometric data, and perform a preliminary model atmosphere analysis of the existing data. Then we discuss in Section 3 additional spectroscopic observations of this object, which are then analyzed in Section 4 using appropriate model atmospheres for this unusual white dwarf. In Section 5 we discuss our findings with respect to the emerging class of magnetic white dwarfs with patchy atmospheres and summarize our findings. 

\section{The Case of SDSS J0910+2105}

SDSS J091016.43+210554.2 (referred to as J0910+2105 from here on) had been classified as an unresolved DA+DB binary by \citet{Genest19b}. Fits obtained using homogeneous DBA model spectra predicted H and He lines that were totally inconsistent with the observed spectrum (see also below). By using a double degenerate model composed of a DA + DB white dwarf, the H lines match up much better (see the online version of their Figure 25.4). While the theoretical He lines are still too deep, this fit is a significant improvement over the single-star model.

\begin{figure}
 \centering
 \includegraphics[angle=-90,width=3.5in, clip=true, trim=0.8in 0.8in 0.3in 0.5in]{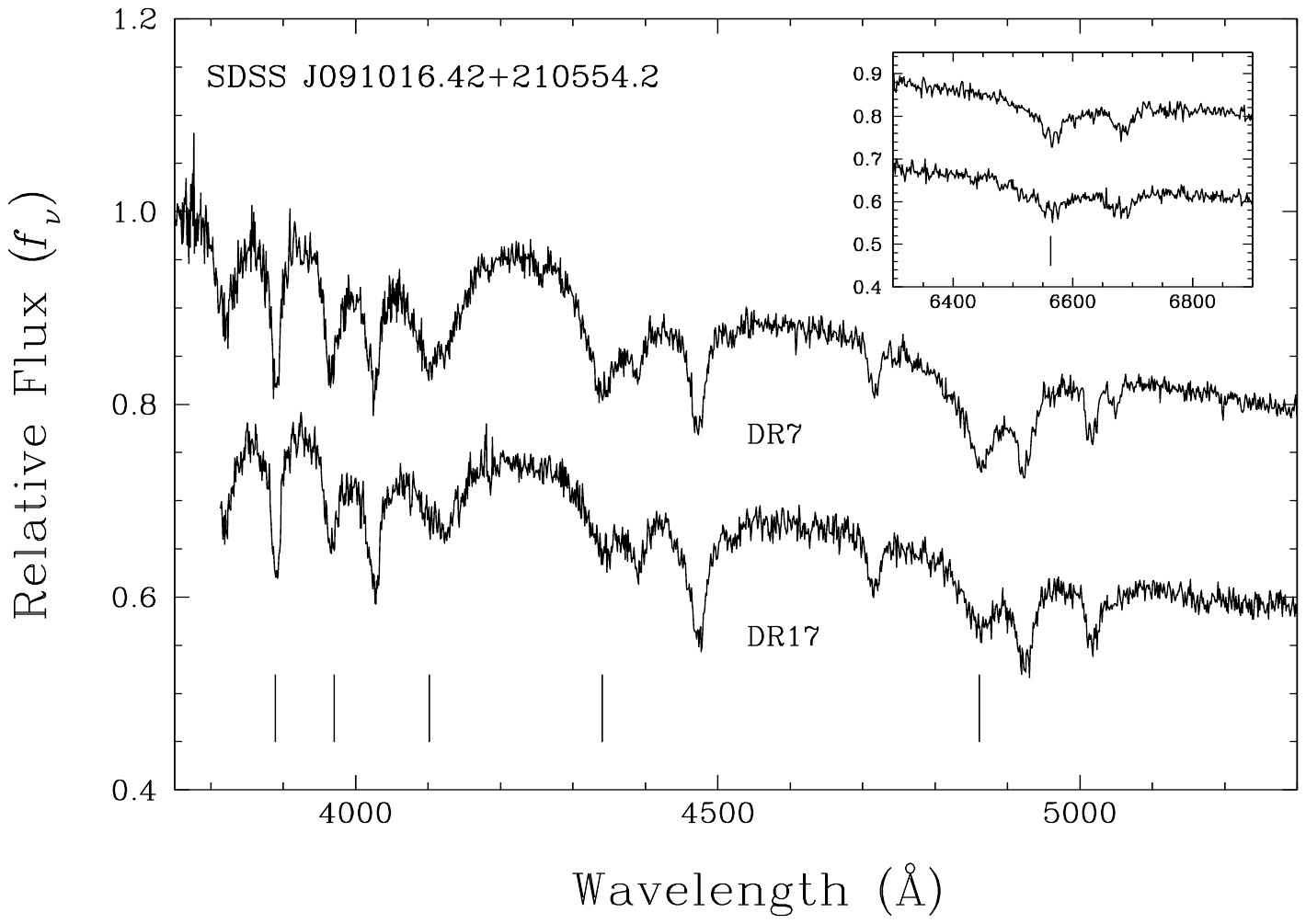}
 \caption{SDSS spectra of J0910+2105. Spectra are offset for clarity. Tick marks indicate the locations of the H lines. The top spectrum shows notably deeper H$\beta$ and H$\gamma$ lines compared to the bottom. The inset shows the H$\alpha$ region where Zeeman-splitting is clearly detected. The exposures are 45 and 48 min long, respectively.}
 \label{fig1}
\end{figure}

Figure \ref{fig1} shows two separate SDSS spectra taken on MJD 53700 and 56003. The inset shows the region around H$\alpha$. We immediately see signs of a magnetic field via Zeeman-splitting in both the H$\alpha$ and \ion{He}{i}  $\lambda$6678 lines. Additionally, the depth of the H lines is noticeably deeper in the top spectrum compared to the bottom, particularly in H$\beta$ and H$\gamma$. An unresolved DA+DB binary system would not produce these spectroscopic variations: the H lines vary in depth but the He lines remain constant. Even if this were a binary, both targets would need to be magnetic given the observed Zeeman-splitting in both sets of lines. 

\begin{figure}
 \centering
 \includegraphics[width=3.5in, clip=true, trim=0.25in 0.75in 0.75in 1.3in]{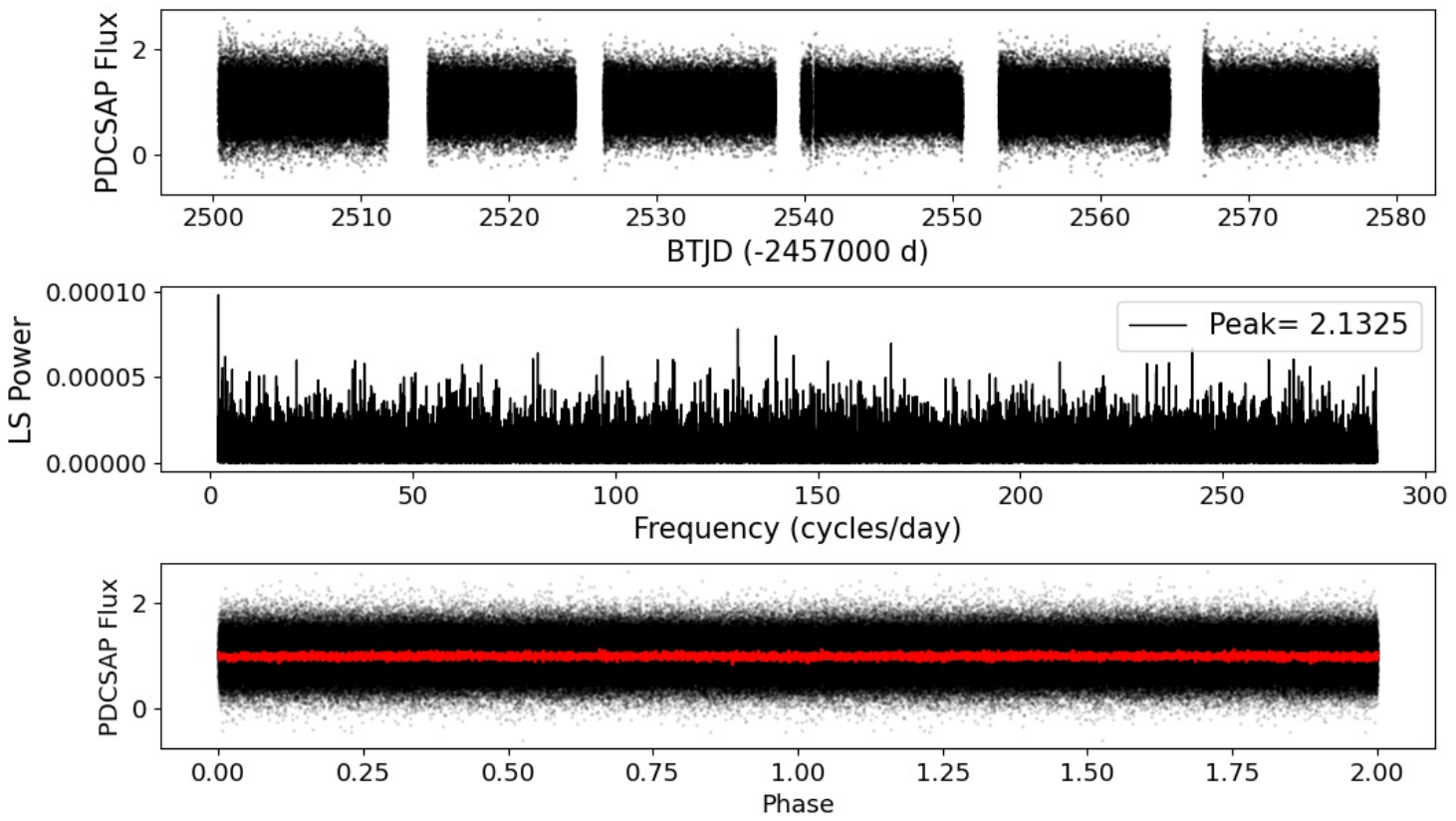}
 \vspace{-1cm}
 \includegraphics[width=3.5in, clip=true, trim=0.25in 0.25in 0.75in 1.3in]{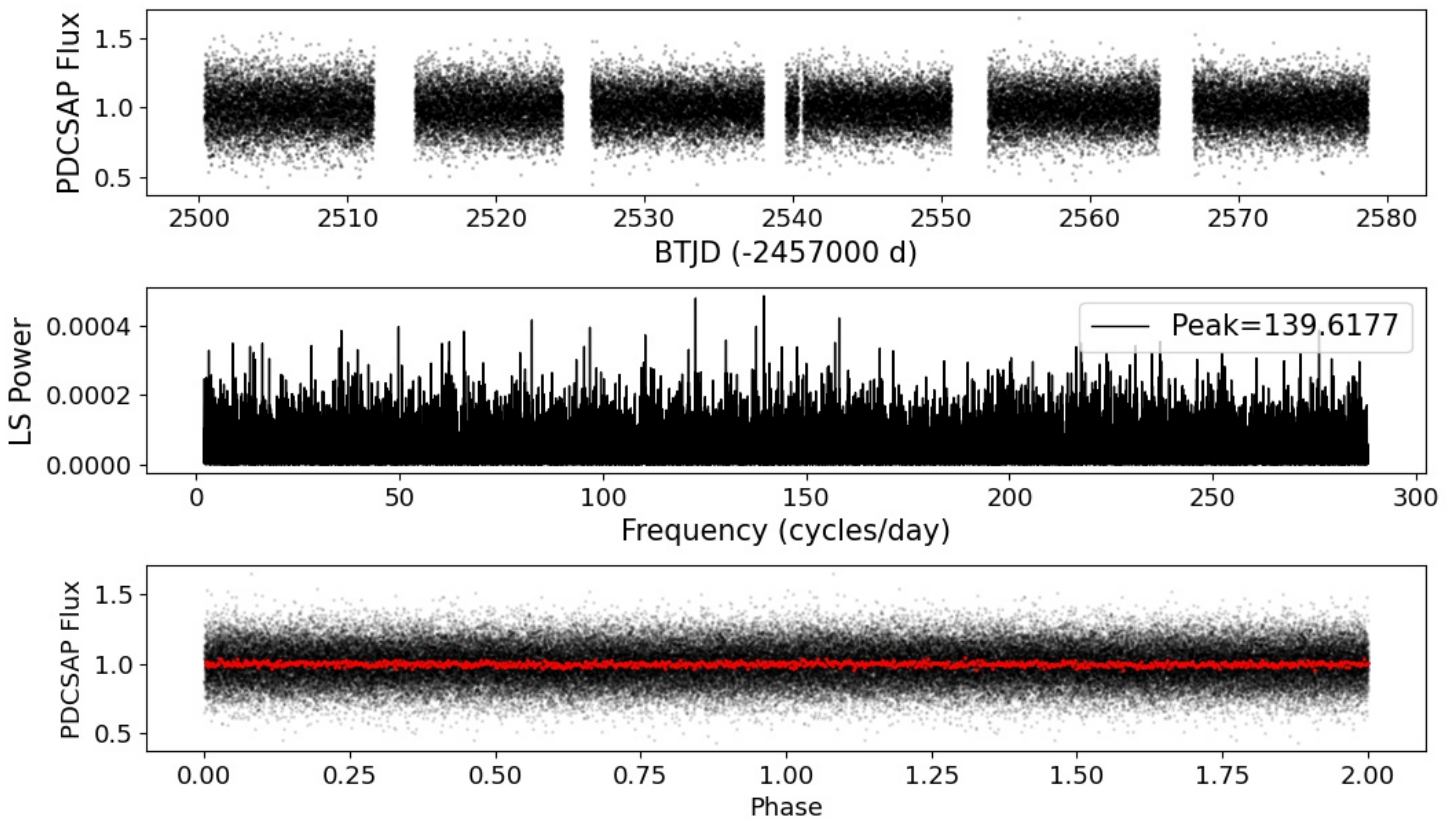}
 \caption{Top: Light curve, Lomb-Scargle periodogram, and phase-folded light curve based on the highest peak in the periodogram for the 20-s cadence data from TESS. Red data points represent the original data binned by 100. Bottom: Similar to the top but for the 2-minute cadence data. Neither cadence shows signs of significant photometric variability.}
 \label{fig2}
\end{figure}

While spectroscopic variations seem likely, photometric variations can provide insight into the rotation period and can be checked using the Transiting Exoplanet Survey Satellite (TESS). J0910+2105 (TIC 85911578) has 20-second and 2-minute cadence data from sectors 44, 45, and 46. We search for frequencies ranging from 5 to 1098 minutes in the 2-minute cadence data, and 1 minute to 1098 minutes for the 20-second cadence. For both sets we used 8 million equally spaced frequency bins. Figure \ref{fig2} shows the results from our TESS search. We do not see signs of periodic photometric variability in either set of data, but this does not rule out the possibility of spectroscopic variations. \citet{Robinson83} showed that GD 323 does not exhibit photometric variations despite having clear spectroscopic variations \citep{Pereira05}.

Table \ref{tab1} shows J0910+2105's parameters from Gaia DR3. In addition, the SEGUE Stellar Parameters Pipeline \citep{Lee08} provides a radial velocity measurement of $61.5 \pm 34.2$ km s$^{-1}$ for J0910+2105 based on the spectrum obtained on MJD 53700. Using the astrometry provided by Gaia DR3 and the radial velocity obtained by the SDSS, we obtain Galactic space velocities of $U = -53 \pm  23, V = -18 \pm 12$, and $W= 29 \pm  22$ km s$^{-1}$. These values are consistent with the disk population \citep{Fuhrmann04}. 

\begin{table}
    \centering
    \caption{\label{tab1} Astrometric and photometric data from Gaia DR3 for J0910+2105.}
    \begin{tabular}{|c|c|} 
    \hline
       Gaia DR3 ID & 637142391019402880 \\
       RA (J2000) & 09:10:16.444 \\
       Dec. (J2000) & +21:05:54.197 \\
       $\pi$ (mas) & 10.381 $\pm$ 0.069 \\
       $\mu_{\alpha}$ (mas/yr) & $-31.883 \pm 0.074$ \\
       $\mu_{\delta}$ (mas/yr) & 7.268 $\pm$ 0.066 \\
       $G$ (mag) & 16.359 $\pm$ 0.003 \\
       $G_{\rm BP}$ (mag) & 16.306 $\pm$ 0.006 \\
       $G_{\rm RP}$ (mag) & 16.463 $\pm$ 0.007 \\
    \hline   
    \end{tabular}   
\end{table}

\begin{figure}
 \centering
 \includegraphics[width=3.4in, clip=true, trim=0.2in 1.7in 0.2in 0.8in]{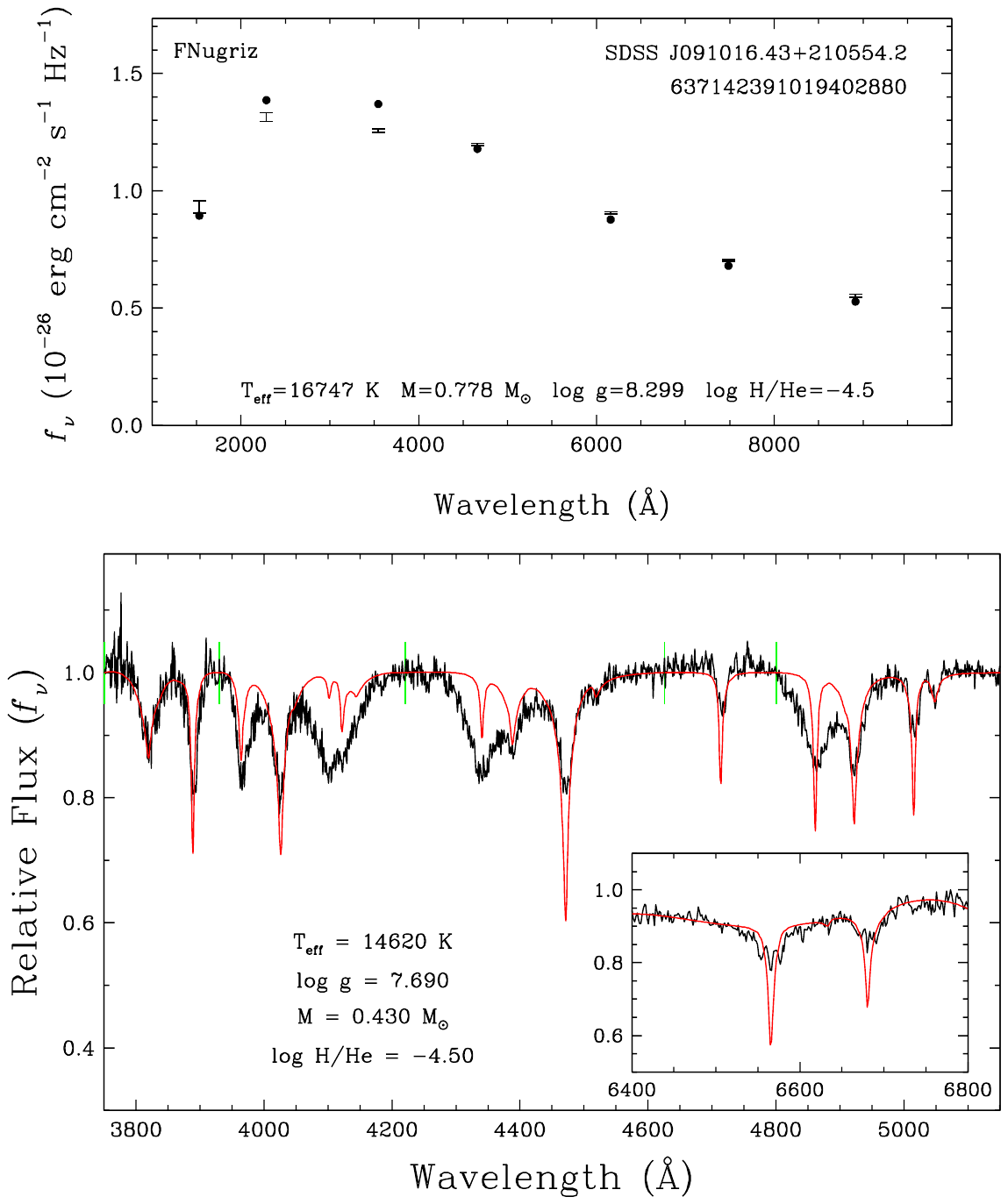}
 \caption{Top: Best photometric fit using GALEX + SDSS + Pan-STARRS  photometry. Bottom: Best spectroscopic fit to the SDSS DR17 spectrum. Both fits assume homogeneous H/He models. The predicted line strengths are either too shallow or too deep for most of the absorption features in the observed spectrum.}
 \label{fig3}
\end{figure}

We use the Gaia DR3 distance, GALEX FUV and NUV, SDSS $u$, and Pan-STARRS $griz$ photometry to obtain a photometric fit of J0910+2105. Figure \ref{fig3} shows our best photometric and spectroscopic fits using homogeneous H/He atmosphere models. While both fits are obtained independently, the H abundance used in the photometric fit is set to the value obtained from the spectroscopic solution. Our photometric solution ($T_{\rm eff} = 16747\pm170$ K, $M=0.778\pm0.011\ M_\odot$, $\log g=8.299\pm0.013$) indicates a fairly cool and massive DBA star. Not only is our spectroscopic solution ($T_{\rm eff} = 14620\pm120$ K, $M=0.430\pm0.055$ $M_\odot$, $\log g=7.690\pm0.135$, $\log {\rm H/He}=-4.50\pm0.09$) completely inconsistent with the photometric solution, the model spectrum is also clearly at odds with the observations, as previously concluded by \citet{Genest19b}. Indeed, the resulting model predicts notably deeper He lines and shallower H lines than what is observed, a discrepancy which is also responsible for the large uncertainties in the derived parameters. Note that this model does not take magnetism into account, however including this would only help with fitting the Zeeman-split components and not significantly affect the overall line strengths. Hence, fits using a homogeneous atmosphere fail to match the spectrum of this target.

Spectroscopic follow-up is necessary to confirm if J0910+2105 is indeed an unresolved double degenerate binary, or if it is a single star with an inhomogeneous atmosphere. In addition, we can analyze several key features of this unique object: the magnetic field strength, the rotation period, and the distribution of H and He in the atmosphere. We present the results from our observations of J0910+2105 in the next section. 

\section{Observations}

We obtained time-series spectroscopy of J0910+2105 using the Apache Point Observatory (APO) 3.5m telescope equipped with the Kitt Peak Ohio State Multi-Object Spectrograph (KOSMOS) over multiple nights. Our first sequence comprised of $42\times5$ min long back-to-back exposures on UT 2023 April 13. We increased the exposure time for our second sequence ($25\times10$ min on UT 2023 Apr 17) after analyzing the first night of data and confirming that the target's rotation period is longer than a few hours. On UT 2023 April 24 we attempted our third sequence, however poor weather conditions restricted us to only two 10 min exposures. Our last set of APO observations were taken on UT 2023 May 26 with $9\times10$ min exposures. All of our APO data were taken using the Blue grism and the 2.1" slit. On UT 2023 April 13 we used the high slit position which provides wavelength coverage from 4150 $-$ 7050{\AA} and a resolution of 1.4 {\AA} per pixel. For the remaining APO observations we used the center slit position which changes the wavelength range to  3800 $-$ 6600{\AA}.

Given the target's hour-scale rotation period, we obtained six additional exposures across four nights (UT 2023 April 22 $-$ 25) using the 6.5m MMT with the Blue Channel Spectrograph. We used the 1.25" slit with the 500 l mm$^{-1}$ grating which yields a resolution of 1.2 {\AA} per pixel. 

\begin{figure}
    \centering
    \includegraphics[width=3.5in,clip=true,trim=0in 0in 0in 0in]{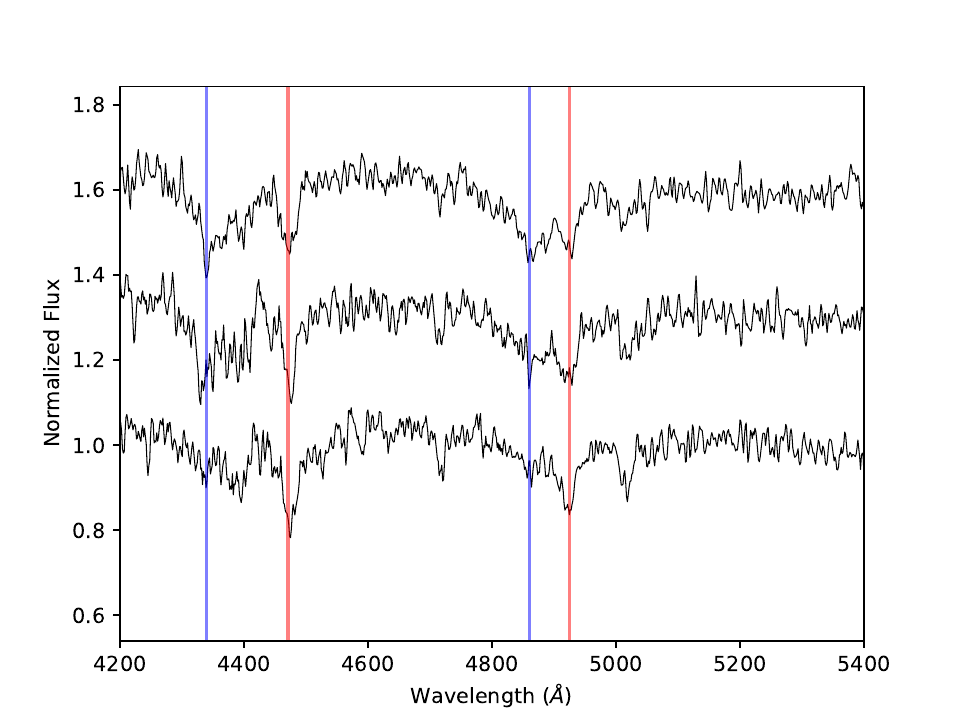}
    \includegraphics[width=3.5in,clip=true,trim=0in 0in 0in 0in]{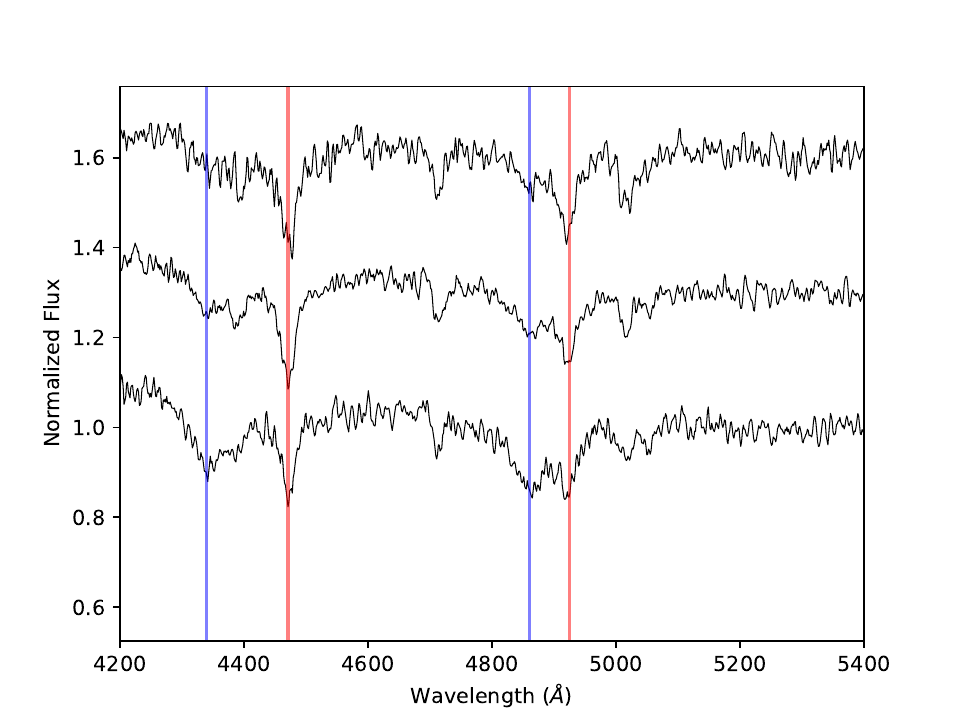}
    \caption{Sample of spectra taken at APO on UT 2023 April 13 (top panel) and UT 2023 April 17 (bottom panel). Spectra are plotted in chronological order from bottom to top. The red lines denote the positions of He absorption lines while blue lines denote H$\gamma$ and H$\beta$. The spectra are offset for clarity and smoothed by 3 pixels for display purposes.}
    \label{fig4}
\end{figure}

We reduced all of our data using the standard IRAF routines. Figure \ref{fig4} shows example exposures from the beginning, middle, and end of the first two nights at APO. All of our spectra are available in the online journal, as well as the UT times at the middle of each exposure. Note that certain exposures are combined. We use the time at the middle of the combined exposure for these. The changes in the H$\beta$ and H$\gamma$ lines are most noticeable in these exposures given the longer baseline. On UT 2023 April 13 (top panels), the H lines are very weak at the beginning of our run. After about seven exposures, the line becomes more pronounced, appearing very clearly by the end of the night at a strength similar to the He lines. Interestingly, the He lines appear constant across all spectra. This suggests the variations in the H$\beta$ line are due to rotation of the white dwarf, with an inhomogeneous distribution of H in the atmosphere. 

Since we did not observe a corresponding decrease in the H$\beta$ line on UT 2023 April 13, it was clear that the rotation period is longer than our initial baseline. Hence we increased our exposure time to 10 minutes to boost the signal-to-noise (S/N) ratio of each spectrum on UT 2023 April 17. Here the H$\beta$ line begins strong (bottom panel in Figure \ref{fig4}), but becomes diminished towards the end of the night. This set of observations confirmed the variations in the H$\beta$ feature are caused by rotation. 

\section{Analysis}

\subsection{The Rotation Period}

Our first goal is to determine the rotation period of J0910+2105 by quantifying the variations seen in the H and He lines across all of our data. To accomplish this, we fit a Gaussian profile to both H$\beta$ and \ion{He}{i}  $\lambda$4922 and calculate the equivalent width of these profiles for each spectrum. We use LMFIT, a version of the Levenberg-Marquardt algorithm adapted for Python \citep{newville14}, to find the best-fit parameters. We then take the ratio of the equivalent widths to measure how the strength of the H lines vary as a function of time compared to the He lines.

\begin{figure}
 \centering
 \includegraphics[width=3.5in,clip=true,trim=0.33in 0in 0.5in 0in]{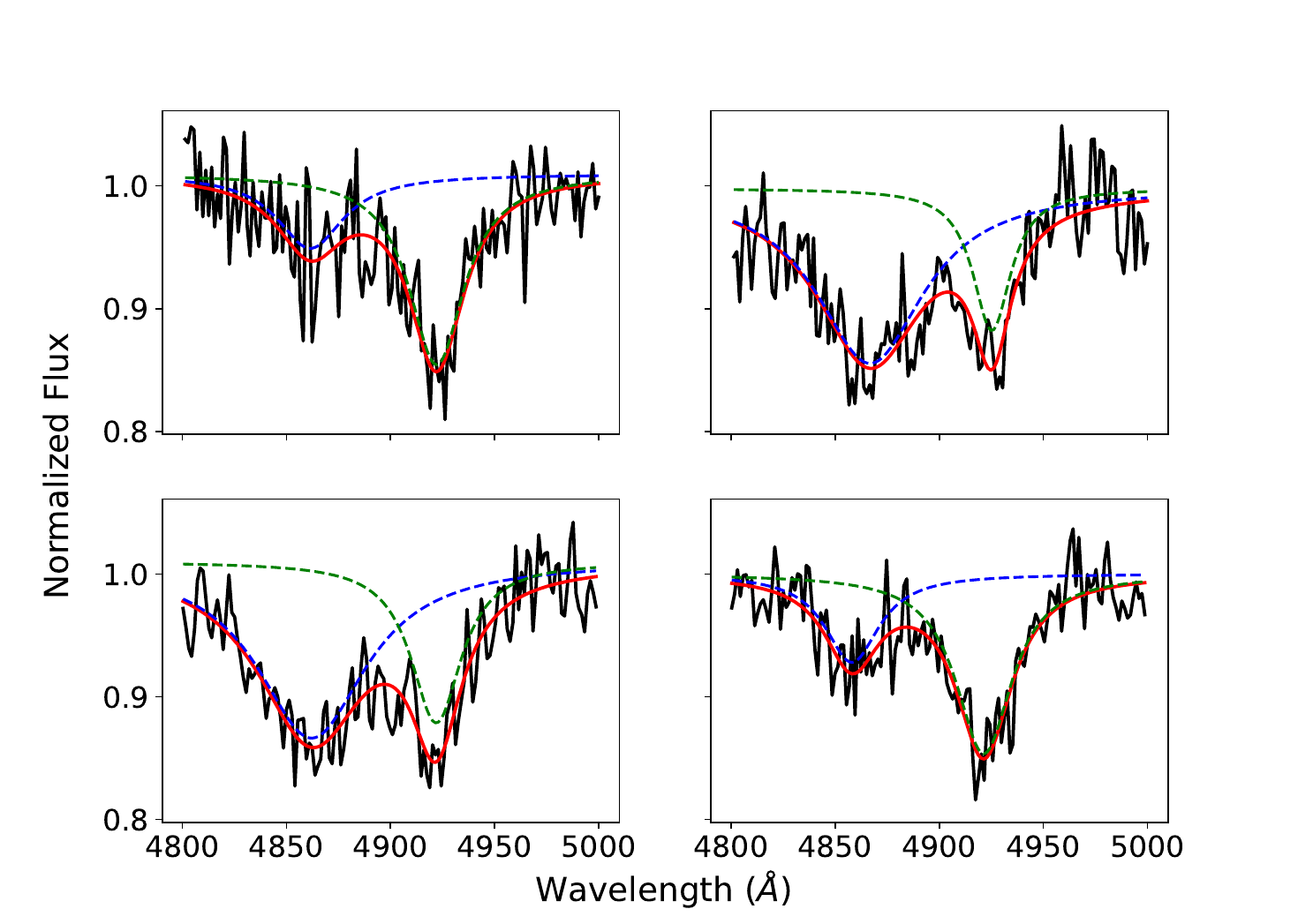}
 \caption{Top: Gaussian fits to H$\beta$ and \ion{He}{i} $\lambda$4922 for the first and last exposure on UT 2023 April 13. The blue line is the best fit Gaussian for H$\beta$ while the green is the best fit for \ion{He}{i}  $\lambda$4922. The red line marks the combined fit of both lines. Bottom: Fits for the second exposure and last exposure on UT 2023 April 17.}
 \label{fig5}
\end{figure}

Figure \ref{fig5} shows example fits to spectra at the start and end of our runs on UT 2023 April 13 and UT 2023 April 17. Here the contrast in the H$\beta$ line is stark, which results in very different equivalent width ratios. In this case, the ratio of the equivalent width of the H$\beta$ line to the \ion{He}{i}  $\lambda$4922 line is 0.45 for the first exposure and 2.96 for the final exposure on UT 2023 April 13. For the UT 2023 April 17 data, the ratios are 1.733 and 0.470 for the second and last exposures respectively. The depth of \ion{He}{i} $\lambda$4922 does not vary significantly across our observations, suggesting a consistent He distribution across the target's surface.  

\begin{figure}
 \centering
 \includegraphics[width=3.4in,clip=true,trim=0.33in 0in 0.5in 0in]{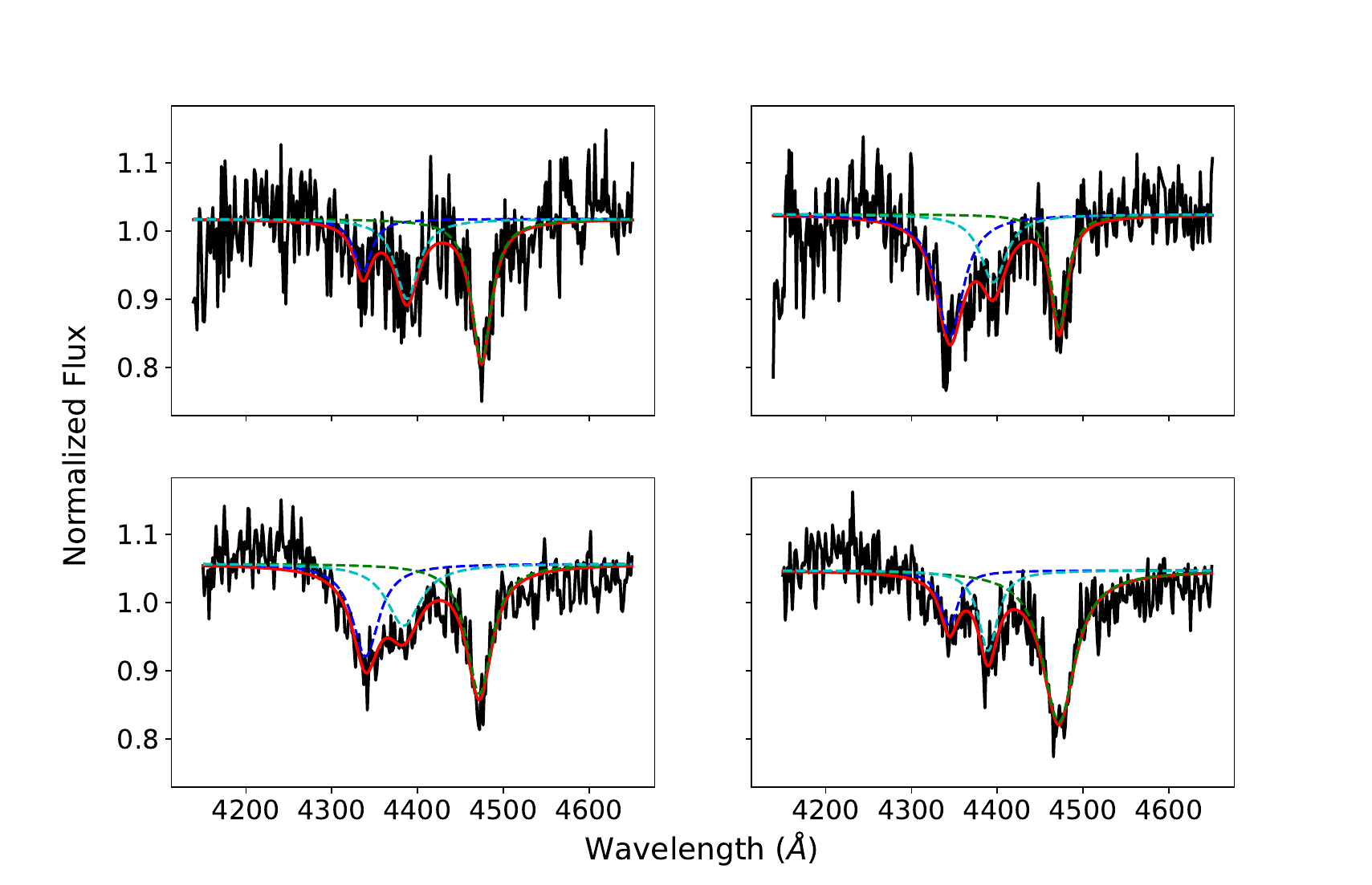}
 \caption{Top: Gaussian fits to H$\gamma$ (blue), \ion{He}{i} $\lambda$4388 (cyan), and \ion{He}{i} $\lambda$4471 (green) for the first and last exposure on UT 2023 April 13. The red line marks the combined fit of all three lines. Bottom: Fits for the second exposure and last exposure on UT 2023 April 17.}
 \label{fig6}
\end{figure}

Assuming these variations in the H$\beta$ line are due to an inhomogeneous distribution of H in the atmosphere, we should expect to see similar variations in other H lines. We perform the same fitting process described above on the H$\gamma$ line, using the nearby \ion{He}{i}  $\lambda$4471  as a reference. However, there is another \ion{He}{i}  line at 4388{\AA} that blends with the H$\gamma$ line. Attempting to fit a single Gaussian to H$\gamma$ produces inferior fits, so we also fit a separate profile to \ion{He}{i}  $\lambda$4388 such that the algorithm clearly distinguishes the two individual profiles and returns a better fit for H$\gamma$. Example fits are shown in Figure \ref{fig6} for the same spectra used in Figure \ref{fig5}.  

\begin{figure}
 \centering
 \includegraphics[width=3.5in,clip=true, trim=0in 0in 0in 0in]{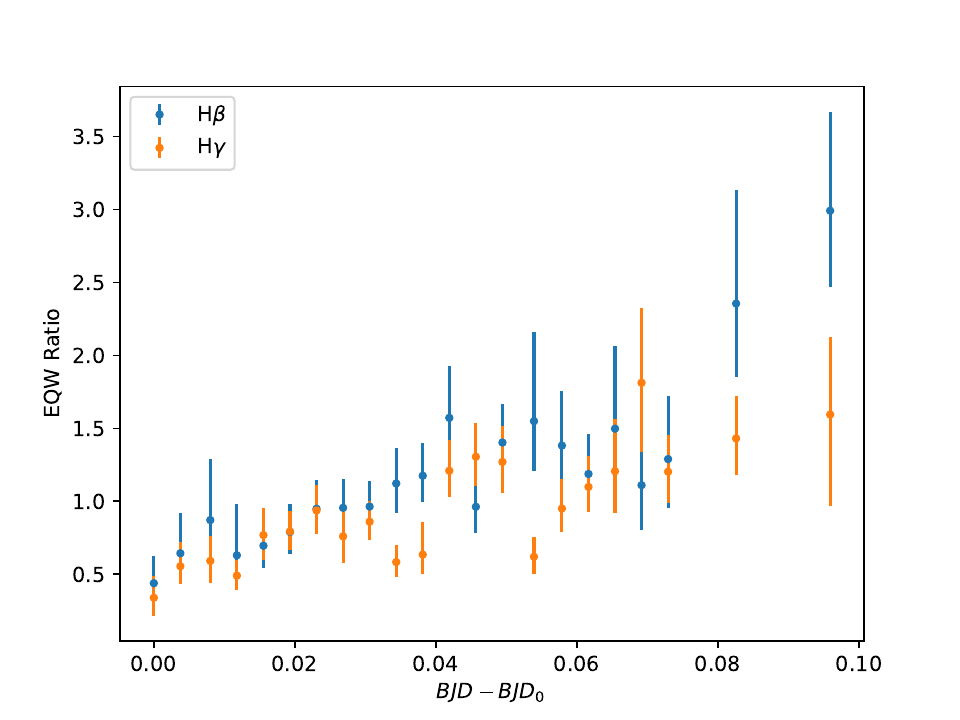}
 \includegraphics[width=3.5in, clip=true, trim=0in 0in 0in 0in]{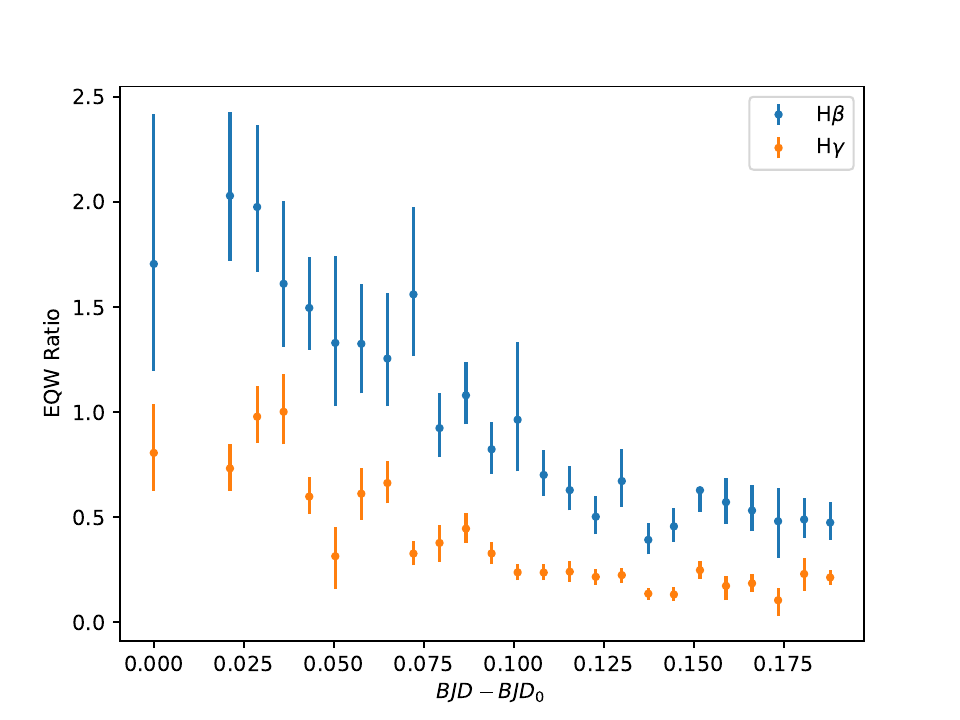}
 \caption{Equivalent width ratios for H$\beta$ and H$\gamma$ from UT 2023 April 13 (top) and 17 (bottom). The H$\gamma$ line tends to be weaker than H$\beta$ due to blending from \ion{He}{i}  $\lambda$4388, so the equivalent width ratios are generally lower. However both lines follow the same trend each night as expected.}
 \label{fig7}
\end{figure}

This fitting method alone however does not allow us to sample the errors in our fits. To obtain an error estimate, we use bootstrapping to repeat the fitting process 10,000 times. We then select the ratios corresponding to 15.9 and 84.1\% in our distribution of 10,000 values as the $\pm 1\sigma$ lower and upper limit. Figure \ref{fig7} shows the ratios and errors for all our spectra from UT 2023 April 13 and 17 for both H$\beta$ and H$\gamma$. On April 13 we clearly see the H lines gradually increase in strength as a function of time. Conversely, on UT 2023 April 17 we see a gradual decrease in the H line strength.

\begin{figure}
 \centering
 \includegraphics[width=3.5in,clip=true, trim=0.7in 0in 0in 0.1in]{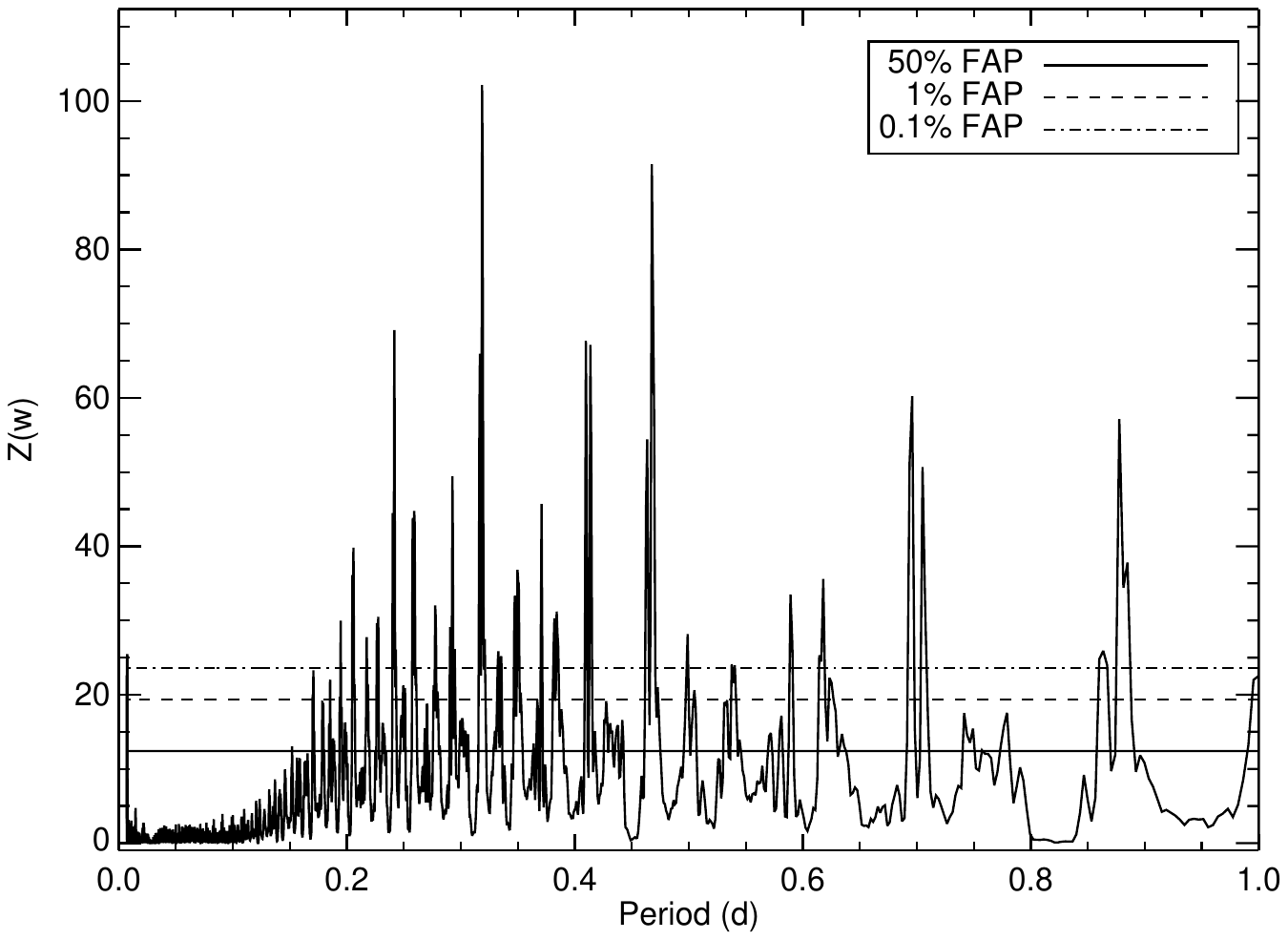}
 \vspace{-0.5cm}
 \includegraphics[width=3.5in, clip=true, trim=0.7in 0in 0in 0.9in]{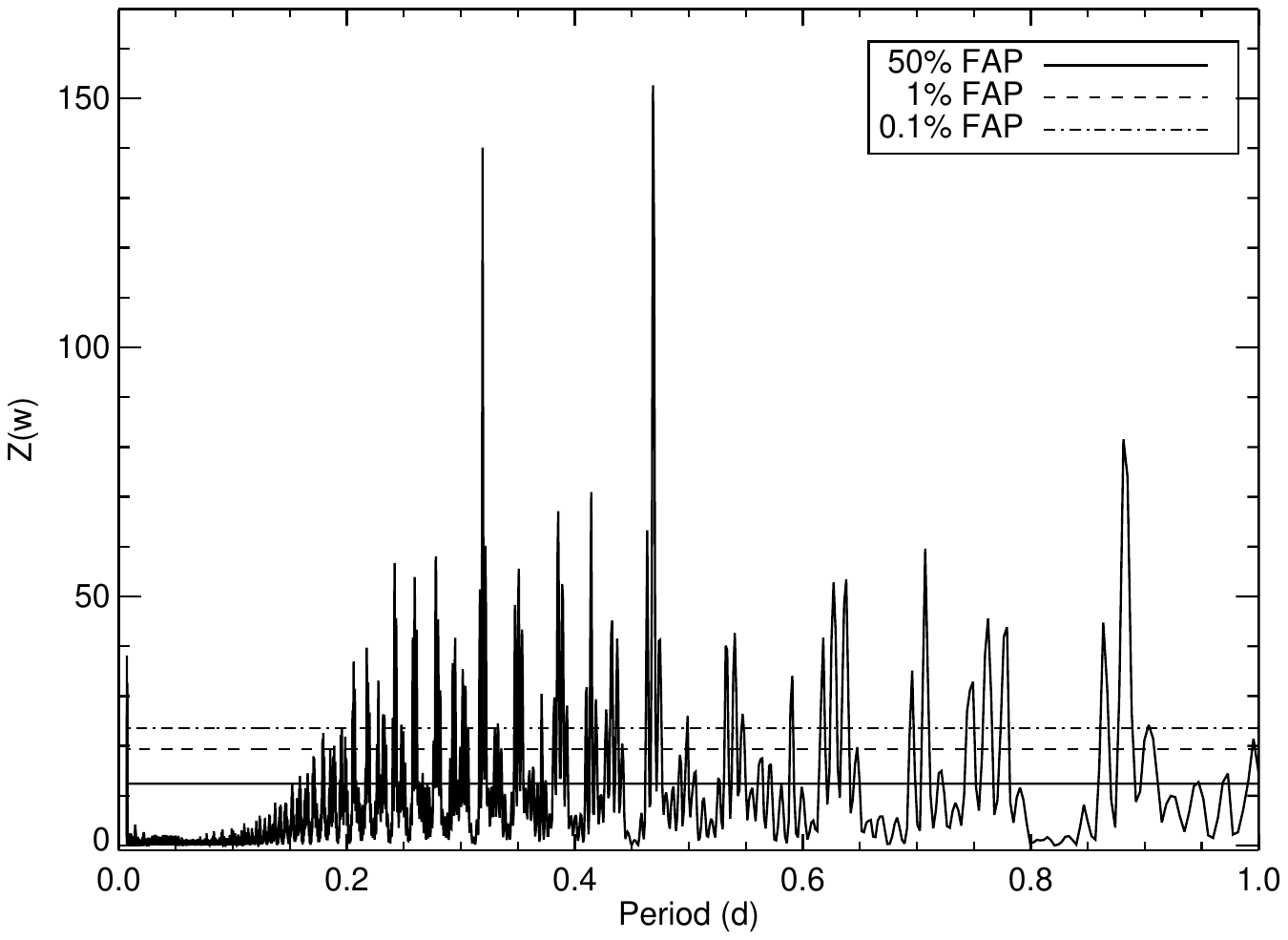}
 \caption{Lomb-Scargle periodograms using the H$\beta$ line (top) and H$\gamma$ line (bottom). Even though significant aliasing is present, current data strongly favor a period of either 7.7 or 11.3 hours.}
 \label{fig8}
\end{figure}

Knowing that the observed changes in the equivalent width ratios are due to rotation of an inhomogeneous atmosphere, we can use the ratios to constrain the target's rotation period. We generate Lomb-Scargle periodograms using the orbital fit code MPRVFIT \citep{DeLee13} in Figure \ref{fig8}. Note that we include the equivalent width ratios from all of the APO and MMT data. We generate two separate periodograms, one for each H line. In both cases, we see two strong signals at 0.319 days (7.66 hours) and 0.470 days (11.28 hours). For the H$\beta$ fits, the signal is slightly stronger at 0.319 days. In contrast, the 0.470 day signal is stronger in the H$\gamma$ fits. Our observations were obtained over a period of up to $\approx4$ hours on a given night, which limits our ability to discriminate between approximately two cycles and three cycles per day. We conclude that the rotation period is either 7.7 or 11.3 hours, though additional observations while the target is visible the entire night would be useful for removing this degeneracy.

\subsection{Modelling the Emergent Fluxes}

We can see from our spectra that the surface composition of J0910+2105 is modulated over time. Here we attempt to develop the theoretical framework to fit these spectra using models that take into account both inhomogeneous H/He surface abundances and the presence of a magnetic field. We discuss each of these physical ingredients in turn.

We first attempt to model the surface abundance inhomogenity by assuming a particular geometry also used by \citet{Beauchamp93} and \citet{Pereira05} in the case of GD 323, which consists of pure H polar caps and a pure He equatorial belt (other geometries are discussed below). We introduce the parameter $\theta_{c}$ to represent the extent of the northern and southern H caps measured from each pole (i.e., $\theta_{c}=0\degree$ means no H caps and thus a pure He atmosphere, while $\theta_{c}=90\degree$ means two full hemispheres and thus a pure H atmosphere). We also denote $\alpha$ as the angle between the magnetic axis and the plane of the sky, as defined in \citet[][$\alpha=90\degree$ means pole-on]{Schnerr2006}.

\begin{figure}
    \centering
    \includegraphics[width=3.5in,clip=true,trim=1.4in 0.5in 1in 1in]{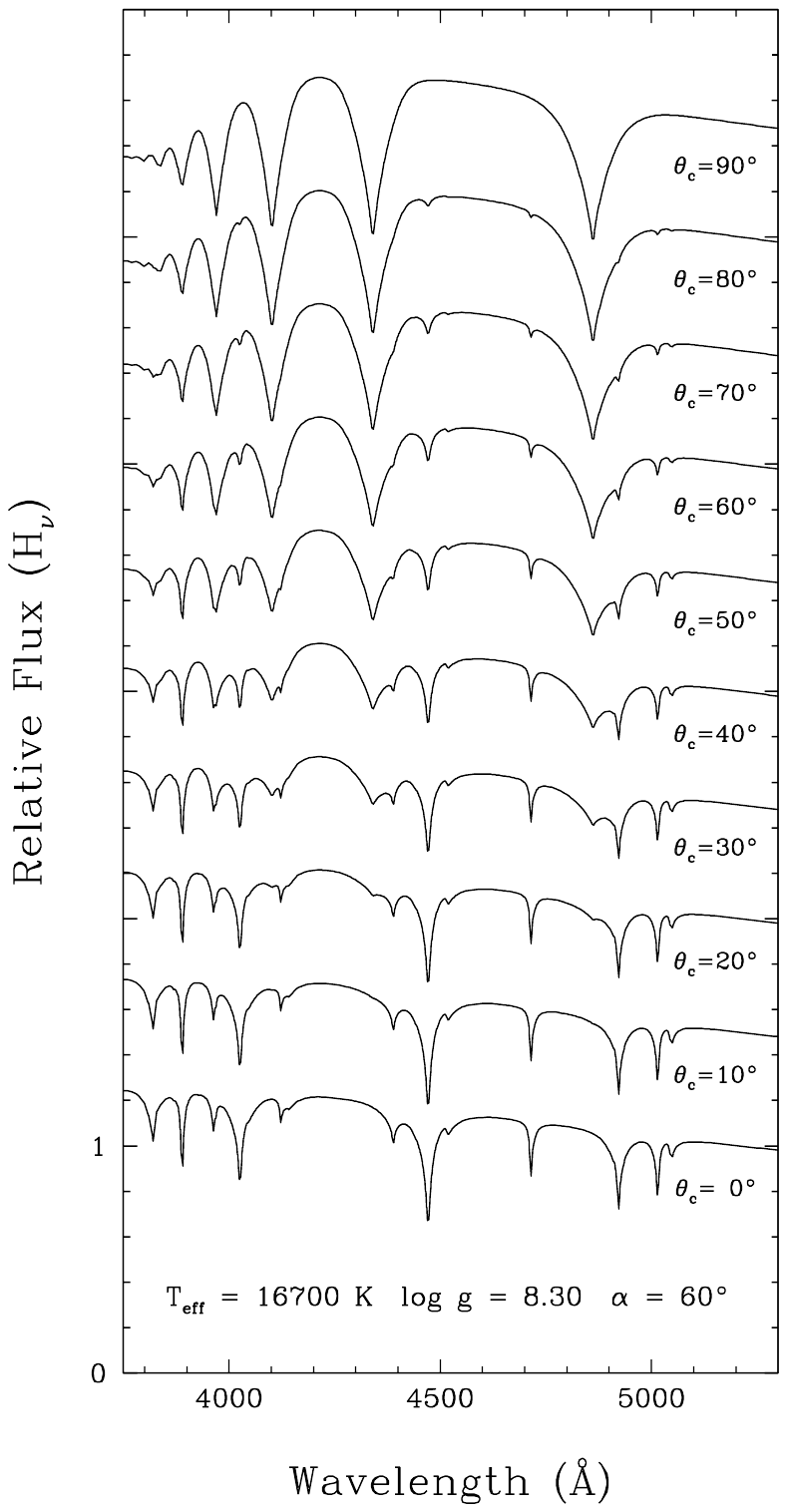}
    \caption{Model spectra (Eddington fluxes) for patchy atmospheres with $T_{\rm eff}=16,700$ K, $\log{g}=8.30$, an angle $\alpha=60\degree$ between the magnetic axis and the plane of the sky, and various values for the extent of the H caps $\theta_{c}$ indicated in the figure. All spectra are normalized to unity at 5200 \AA\ and offset from each other for clarity.}
    \label{fig9}
\end{figure}

The emergent Eddington flux $H_\nu$ is then calculated by numerically integrating the specific intensity $I_\nu$ over the visible surface of the disk given the adopted geometry and viewing angle. We adopt pure H and pure He model atmospheres based on the surface composition of each given element. To do so, we fix the $T_{\rm eff}$ and $\log{g}$ values of both sets of models to the values obtained from the photometric fit, which is certainly a good approximation of the overall physical properties of this star. However, because these parameters have been obtained using homogeneous H/He model atmospheres, we allow them to vary if necessary (our results below indicate that these are appropriate). Sample spectra for these so-called patchy atmospheres are displayed in Figure \ref{fig9} for various values of $\theta_c$, going from a pure He spectrum ($\theta_c=0\degree$) to a pure H spectrum ($\theta_c=90\degree$). The values in between illustrate how the relative strength between H and He lines varies as a function of the extent of the H caps.

Given the adopted atmospheric parameters based on our photometric fit, $T_{\rm eff}=16,700$ K and $\log{g}=8.30$, our pure H model is fully radiative while our pure He model has a well-developed convective zone. Since it is possible that the presence of the magnetic field inhibits convection, at least partially, we also consider below pure He models with fully radiative atmospheres.

To model the magnetic field of the H and neutral He lines, we use a theoretical approach similar to that described in \citet{Bergeron1992} where the total line opacity is calculated as the sum of the individual Stark-broadened Zeeman components. Here, the line displacements and oscillator strengths of the Zeeman components of H$\alpha$ through H$\delta$ are taken from the tables of \citet{Kemic74}, while for the neutral He lines, we use the detailed calculations from \citet{Becken99}, kindly provided to us by S.~Jordan, which have been used successfully in the context of magnetic DB white dwarfs by \citet{Hardy23}. For both H and He lines, the total line opacity is normalized to that resulting from the zero-field solution. The specific intensities at the surface, $I(\nu,\mu,\tau_{\nu}=0)$, are obtained by solving the radiative transfer equation for various field strengths and values of $\mu$ ($\mu= \cos \theta$, where $\theta$ is the angle between the angle of propagation of light and the normal to the surface of the star). In doing so, the polarization of the radiation field is neglected as we are mainly interested in the total monochromatic intensity. Also, the thermodynamic structures are those obtained from our non-magnetic model atmospheres. Because there is no theory that currently treats magnetic and Stark broadening simultaneously, we make the simple assumption that each magnetic Zeeman component is individually Stark-broadened using an approximate treatment for both H and He lines, instead of relying on the detailed line profile calculations of \citet{Tremblay09} and \citet{Beauchamp97} for the H and He lines, respectively. In particular, for the H lines, we rely on the Stark treatment from \citet{Kurucz1970}, which is based on Griem's broadening theory \citep{Griem1960,Griem1967}, as described in \citet{Wesemael1980}, while for the neutral He lines, we used
the electron impact shifts and widths and ion broadening parameters
taken from \citet{Griem1974}, as described in \citet{Wesemael1981}. As a test of our overall procedure, we compared our synthetic spectrum calculations with those of \citet{Hardy2023a,Hardy23} for magnetic DA and DB white dwarfs, respectively, and found an excellent agreement with their models.

\begin{figure}
 \centering
 \includegraphics[angle=-90, width=3.5in, clip=true, trim= 0.8in 0.2in 0.3in 0in]{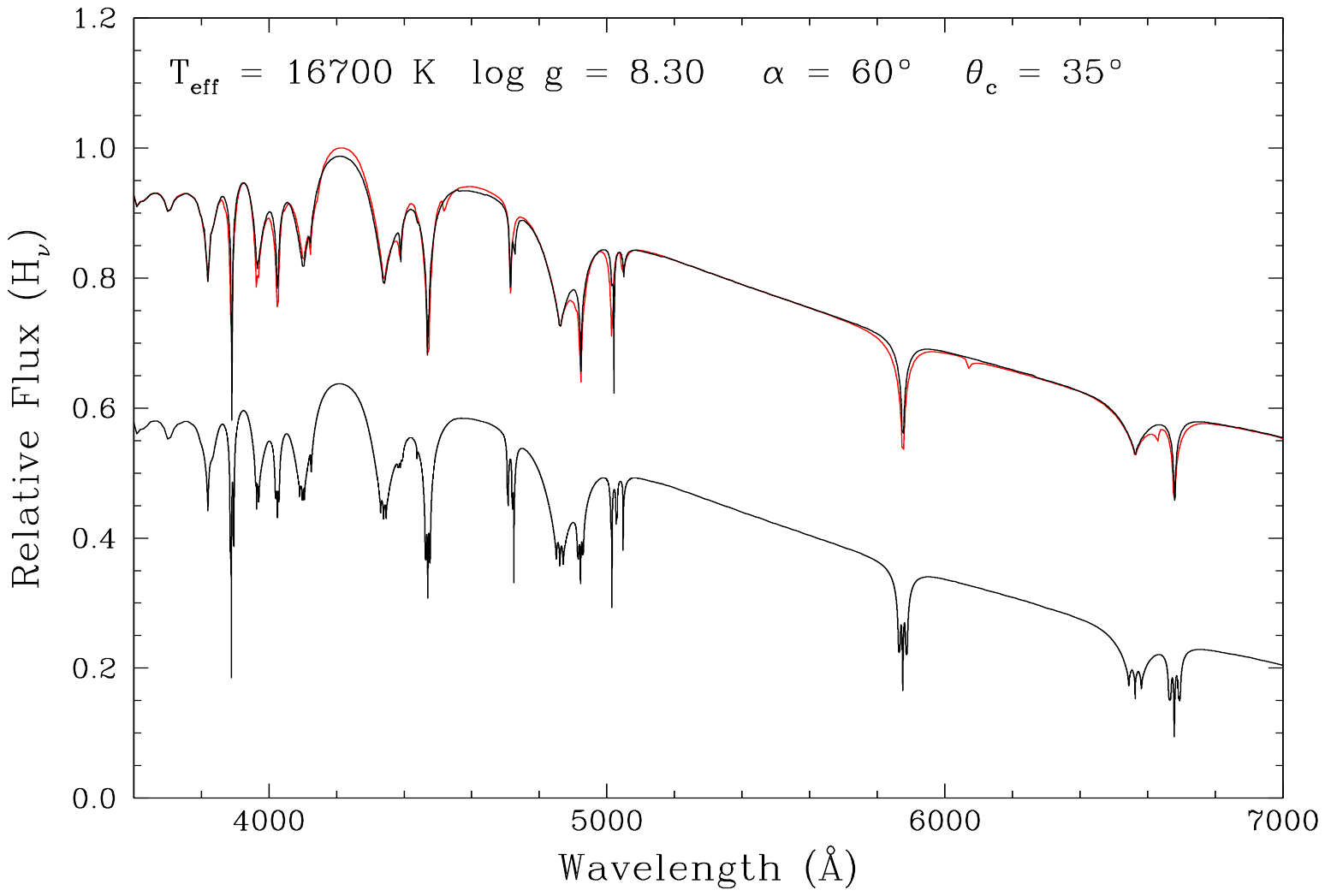}
 \caption{Theoretical spectra obtained from model atmospheres with inhomogeneous surface abundances (see text). The top two spectra compare non-magnetic models using detailed (red) and approximate (black) Stark profiles for both the H and He lines. The bottom spectrum (offset for clarity) includes a magnetic field with a dipole field strength of $B_d=1$ MG.}
 \label{fig10}
\end{figure}

Results from these theoretical calculations are illustrated in Figure \ref{fig10} for H polar caps with $\theta_c=35\degree$ (thus a He equatorial belt $110\degree$ thick), an angle $\alpha=60\degree$ between the magnetic axis and the plane of the sky, and the adopted atmospheric parameters taken from the photometric solution. The top spectra compare two non-magnetic models where H and He lines have been treated with detailed Stark profiles (red spectrum), and with our more approximate Stark profiles (black spectrum); we can see the absence of the \ion{He}{i} forbidden components with this more approximate treatment. Overall, the comparison between these spectra is excellent, and certainly sufficient for our purposes. The bottom spectrum in Figure \ref{fig10} shows the inclusion of a magnetic field in our calculations with a dipole field strength of $B_d=1$ MG; here and in the remainder of our analysis, we simply assume a centered dipole model. The splitting of both the H and He lines can be clearly observed in this magnetic spectrum. We conclude that this theoretical framework of magnetic white dwarfs with inhomogeneous surface abundances is perfectly suited for our spectroscopic analysis, which we now consider.

\begin{figure}
 \centering
 \includegraphics[width=3.5in,clip=true,trim=0in 3.3in 0in 3.3in]{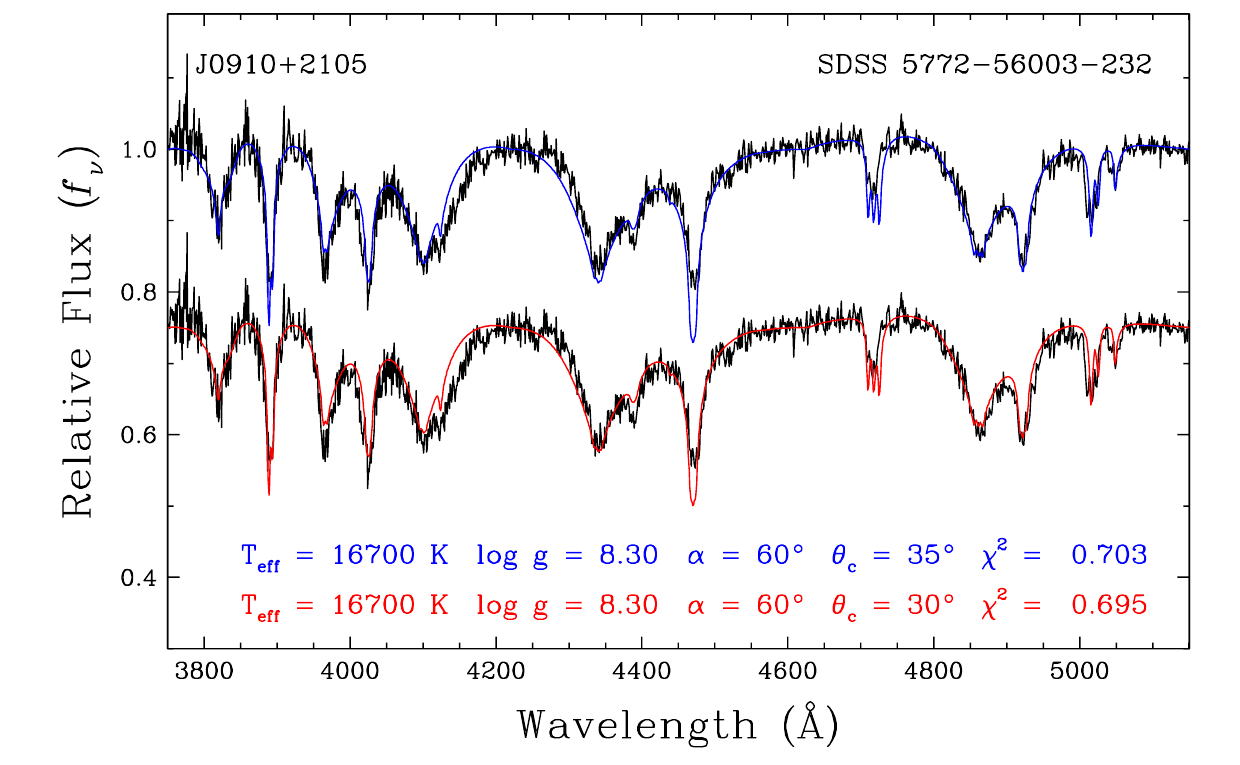} 
 \includegraphics[angle=-90,width=3.3in,clip=true,trim=0.9in 0.7in 0in 1.7in]{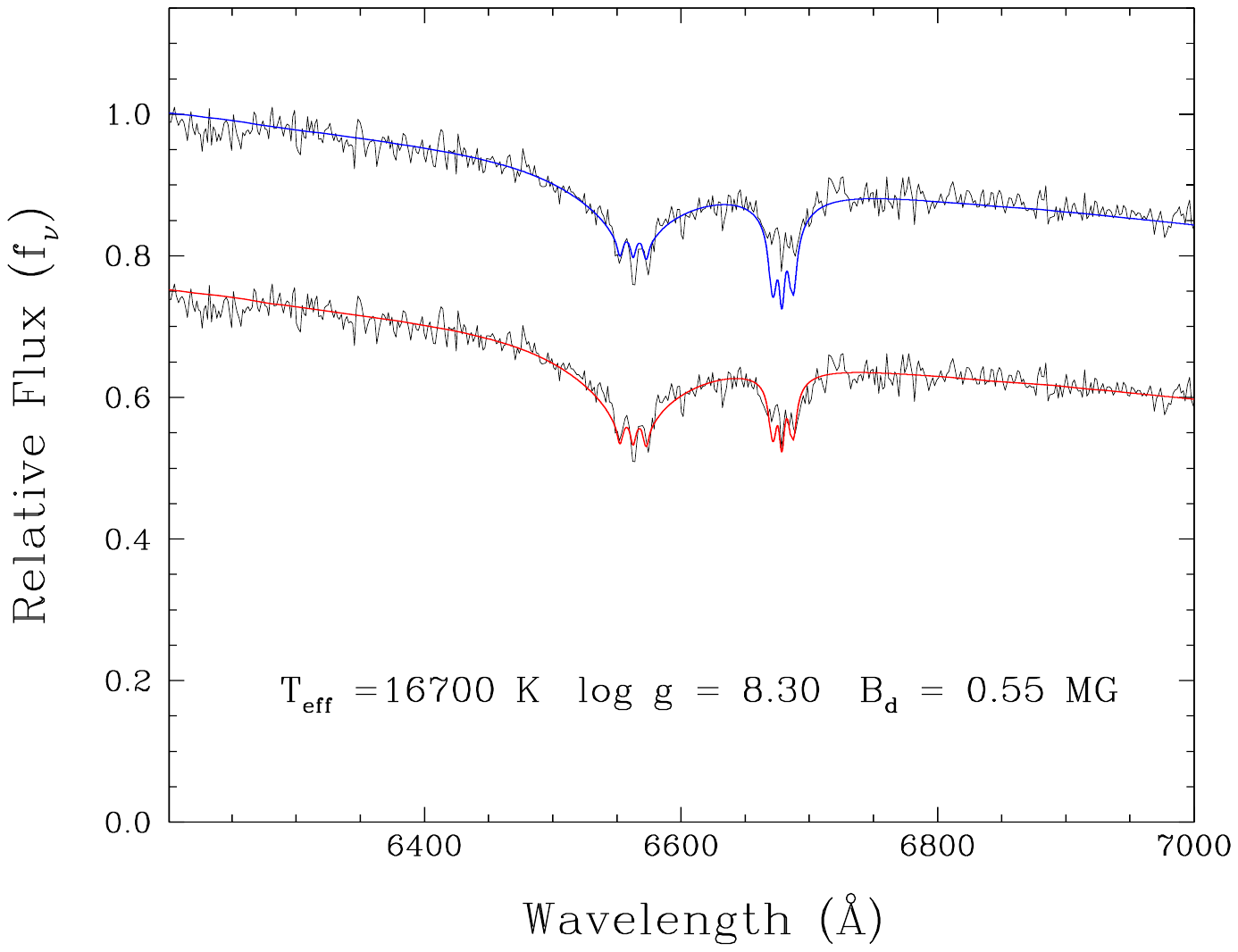}
 \caption{Top: Best-fitting convective (blue) and radiative (red) atmosphere models to an SDSS spectrum of J0910+2105 using inhomogeneous model atmospheres with H polar caps. The size of the H caps $\theta_c$, the angle between the magnetic axis and the plane of the sky $\alpha$, and the $\chi^2$ value of the fit are indicated in the figure. Bottom: Corresponding fits to H$\alpha$ and \ion{He}{i} $\lambda$6678 lines. This portion of the spectrum is used to determine the strength of the magnetic field. Here we find that a centered dipole model with a dipole field strength of $B_d=0.55$ MG provides the best fit; we also find that the He radiative atmosphere produces a deeper line profile for H and a shallower one for He, yielding a better fit overall.}
 \label{fig11}
\end{figure}

Figure \ref{fig11} shows various fits to one of the SDSS spectra, where we emphasize the blue and red regions of the spectrum. We fit the blue spectrum by considering both the size of the H polar caps $\theta_c$ and the angle $\alpha$ as free parameters,  while we use the red region near H$\alpha$ to determine the strength of the magnetic field given that the line splitting is more important there. Our analysis indicates that a centered dipole model with a dipole field strength of $B_d=0.55$ MG provides an excellent description of the observed line splitting, which is the value we use throughout our analysis. Because we expect that a field strength of $B_d\sim0.5$ MG is sufficient to inhibit surface convection based on the findings of \citet{Tremblay15}, who showed that even a field strength of only $\sim$50 kG can suppress convection, we explore both possibilities in our fits.

Our best fits to the blue spectrum using both radiative and convective atmospheres for the He models are displayed in the top panel of Figure \ref{fig11} (the model spectra are properly convolved with the instrumental profile). A comparison of our best fits with that displayed in Figure \ref{fig3} under the assumption of homogeneous H/He atmospheres indicates that our theoretical framework provides a significant improvement and an excellent match to the observations, giving us confidence in our overall approach. Note that with the exception of the line centers where Zeeman splitting is observed, the improvement in the quality of the fits originates mostly from the treatment of patchy atmospheres rather than the inclusion of the magnetic field. 

Even though the fit to the blue spectrum using radiative He model (shown in red) provides a small improvement over that obtained with convective model (shown in blue), as measured by the smaller $\chi^2$ value but also by the overall quality of the fits, the greatest improvement in the radiative fit is for \ion{He}{i} $\lambda$6678 shown in the bottom panel of Figure \ref{fig11} where we contrast again the convective and radiative solutions. We believe that these results represent a strong indication that the small $\sim$0.5 MG magnetic field present in J0910+2105 is indeed sufficient to suppress convective energy transport, at least in the atmospheric regions of this star. Consequently, we focus on the radiative fits for the remainder of the analysis. 

Note that although our radiative fit of the \ion{He}{i} $\lambda$4471 line is slightly better than the convective fit, the observed line remains noticeably shallower, which may indicate the limitations of our overall theoretical framework, in particular given the fact that the line core of this particular line is formed high in the atmosphere.

We used this theoretical framework to fit all the spectroscopic observations presented in Section 3, the results of which are discussed in the next subsection. Fits to all of our spectra are presented in the online journal. Although it was possible to reproduce successfully all our spectra using our assumed geometry of H polar caps and He belts, we also explored other surface geometries. For instance, we reversed the abundances and assumed He caps with an equatorial H region. We also assumed a single H (or He) region defined between two predefined polar angles that are not necessarily symmetrical around the equator, as opposed to our previous assumption. This particular geometry could be used to model a white dwarf such as Janus \citep{Caiazzo23} with pure H and pure He hemispheres, for instance. Although we do not present the results of these calculations here, we found that we could always achieve good fits to the observations using any of these assumptions. Note that it is always possible that our assumption of H caps and a He equatorial region may not be representative of the true distribution of H and He of the surface of the stellar disk, but instead only serves as a proxy of the surface area covered by each element. We test our adopted geometry further in the next section using the so-called oblique rotator model.

\subsection{Model Fits to the Time-Resolved Spectra}

A physical model to explain the observed abundance variations in J0910+2105 is the oblique rotator model \citep{Stibbs50,Monaghan73}, where the magnetic axis is tilted with respect to the rotation axis. Here, because the two axes are misaligned, the observer would see a different surface abundance (and magnetic field) distribution across the stellar surface due to rotation. The oblique rotator model denotes $\beta$ as the angle between the two axes, and $i$ as the angle between the line of sight and rotation axis (see Figure 2 of \citealt{Bailey11} for an image of the geometry of this model). This model has been successfully used in the literature to constrain the magnetic field geometry of Ap stars and magnetic white dwarfs based on spectropolarimetry \citep[e.g.][]{Liebert77,Borra78,Bailey11}. For example, \citet{Valyavin05} analyzed the magnetic white dwarf, WD 0009+501, and found a relatively large $\beta = 117\pm24\degree$ and a viewing angle of $i = 63\pm24\degree$ to explain the spectral variations in that object.

Our optical spectroscopy data on J0910+2105 do not have sufficient signal-to-noise ratio to constrain the change in the average surface
magnetic field strength over time. Nevertheless, the observed variations in the H line strengths can be used to constrain the system geometry using the oblique rotator model. Since the lowest $\chi^2$ solutions are typically found for a cap size of $35\degree$, we fix $\theta_c = 35\degree$ for the following analysis. 

\begin{figure*}
    \centering
    \includegraphics[width=3.5in,clip=true,trim=1in 3.3in 0in 2.75in]{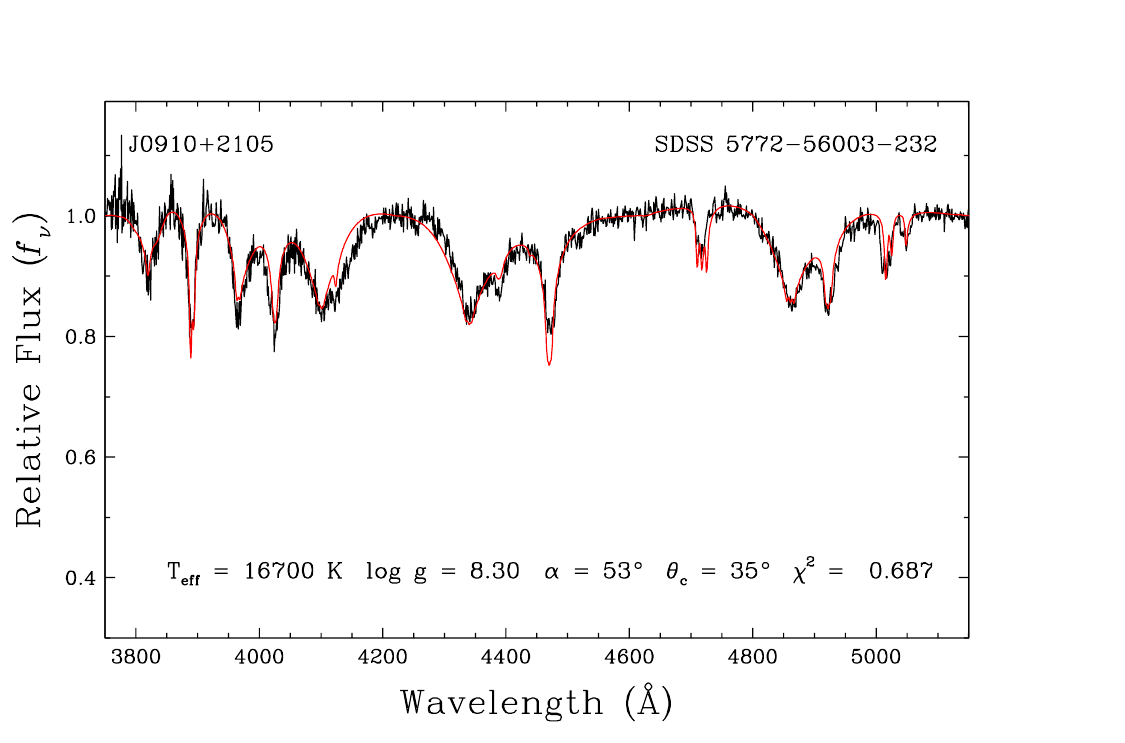}
    \includegraphics[width=3.5in,clip=true,trim=1in 3.3in 0in 2.75in]{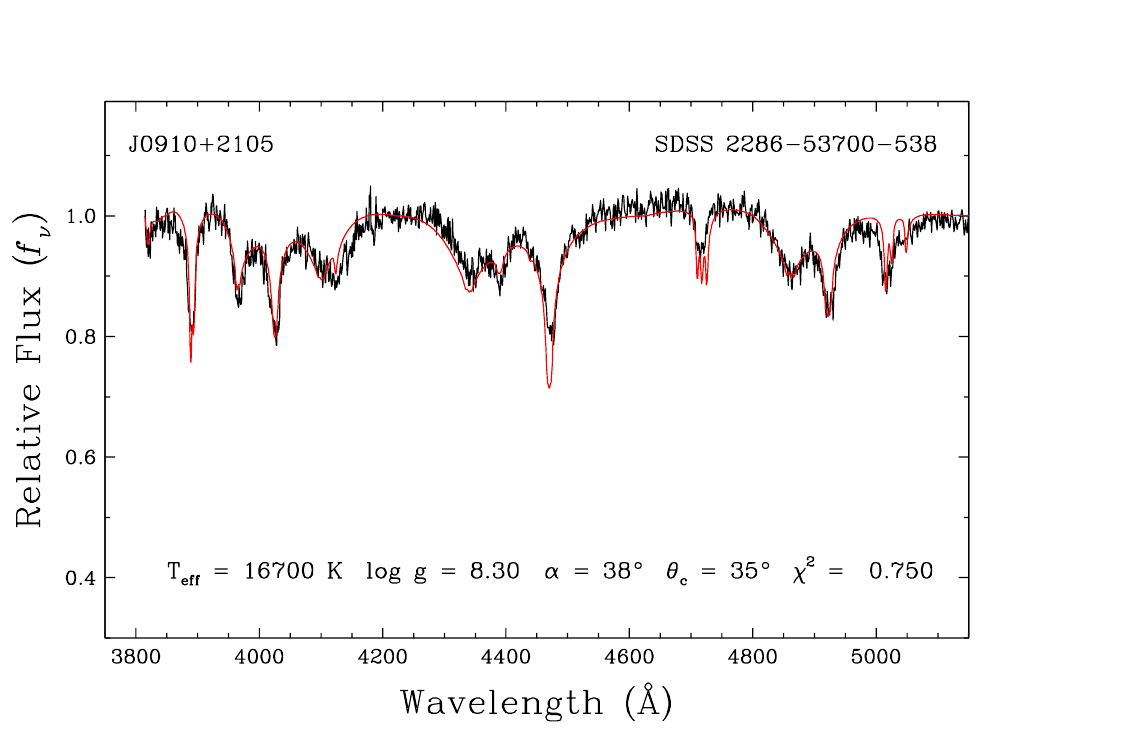}
    \includegraphics[width=3.5in,clip=true,trim=1in 3.3in 0in 2.75in]{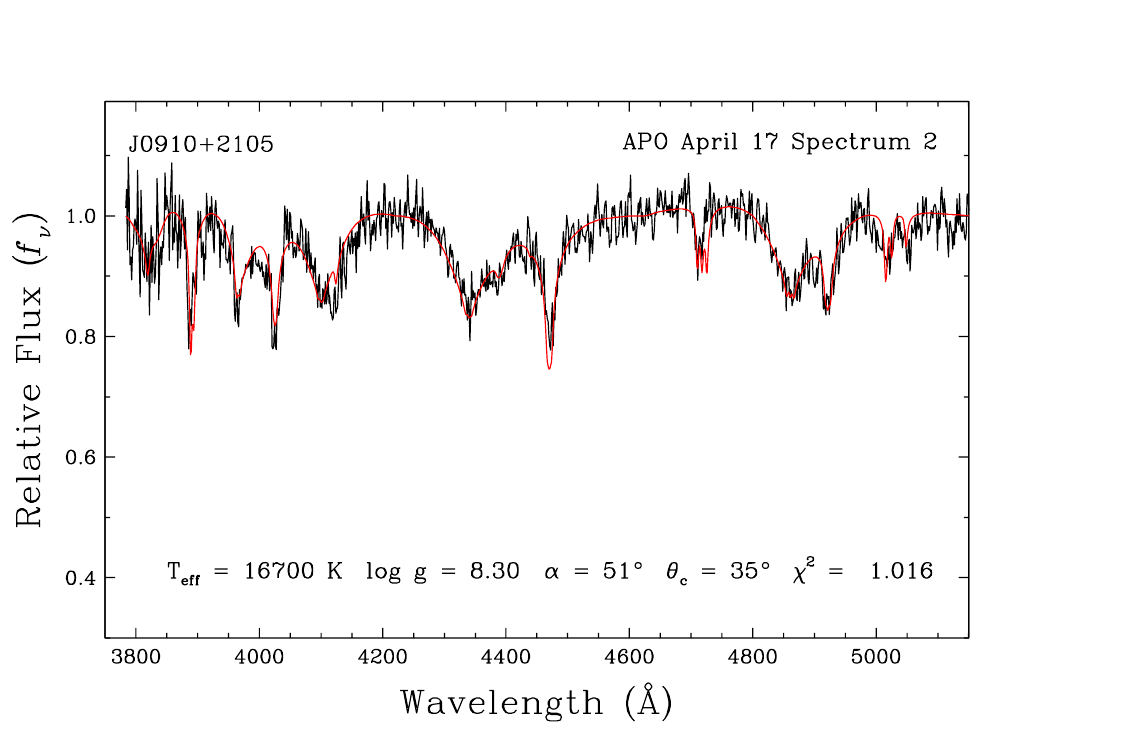}
    \includegraphics[width=3.5in,clip=true,trim=1in 3.3in 0in 2.75in]{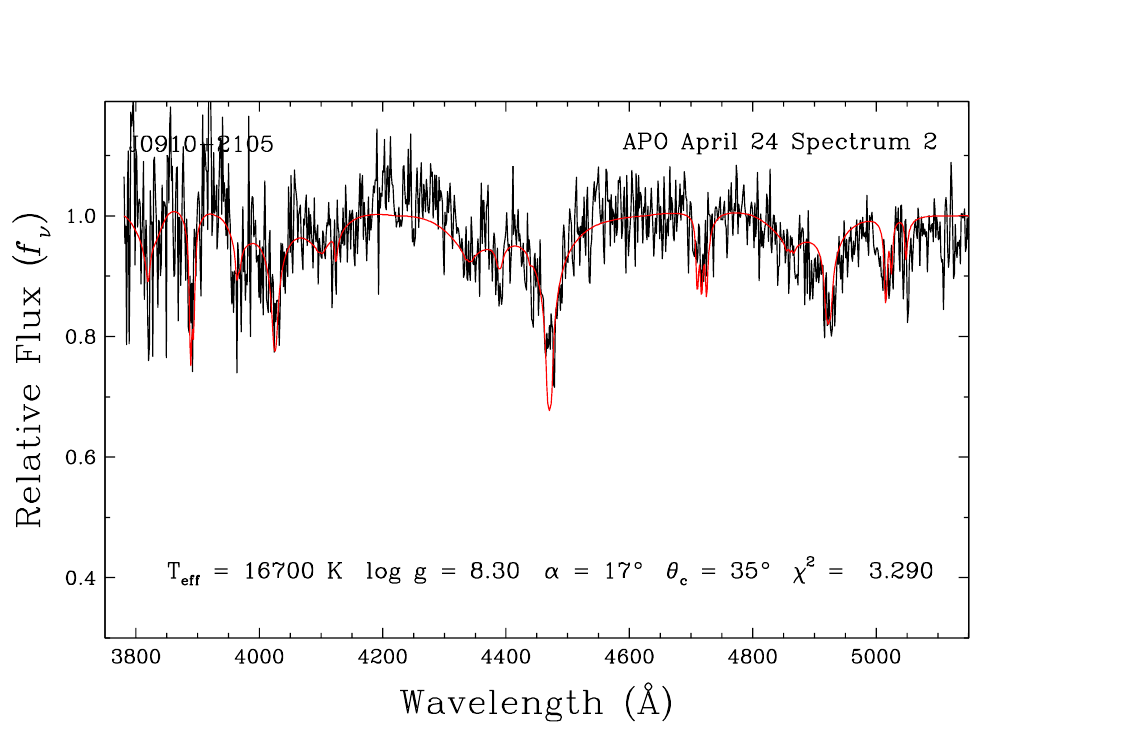}
    \includegraphics[width=3.5in,clip=true,trim=1in 3.3in 0in 2.75in]{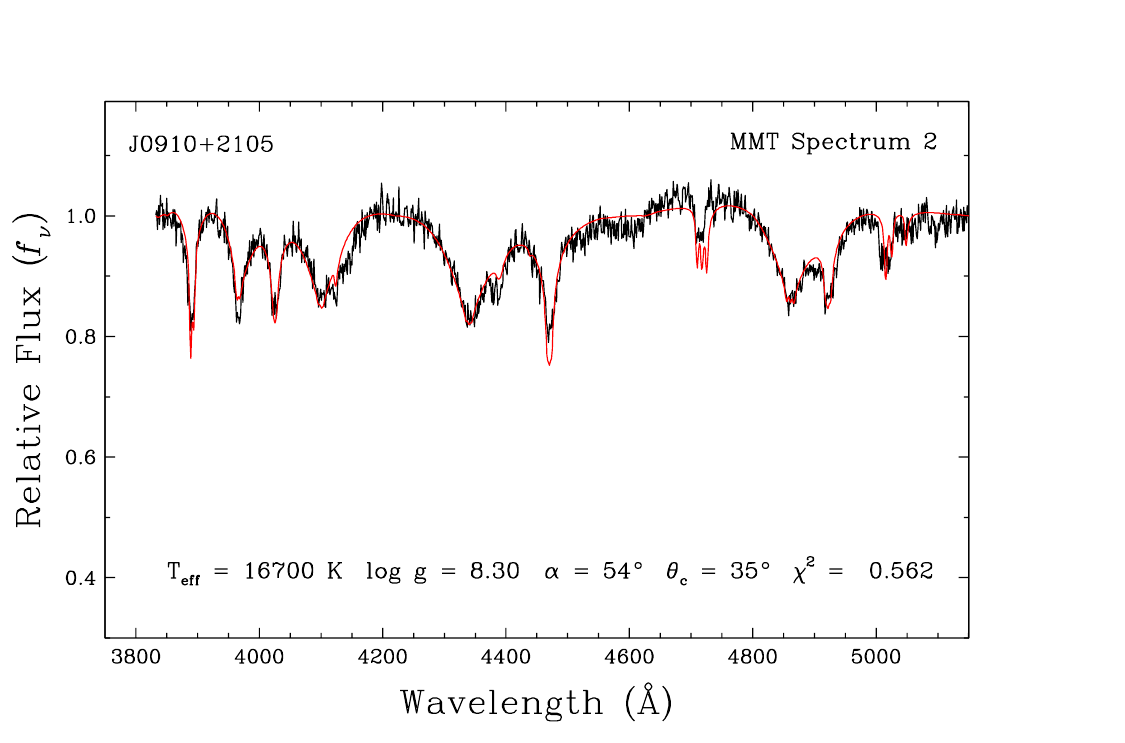}
    \includegraphics[width=3.5in,clip=true,trim=1in 3.3in 0in 2.75in]{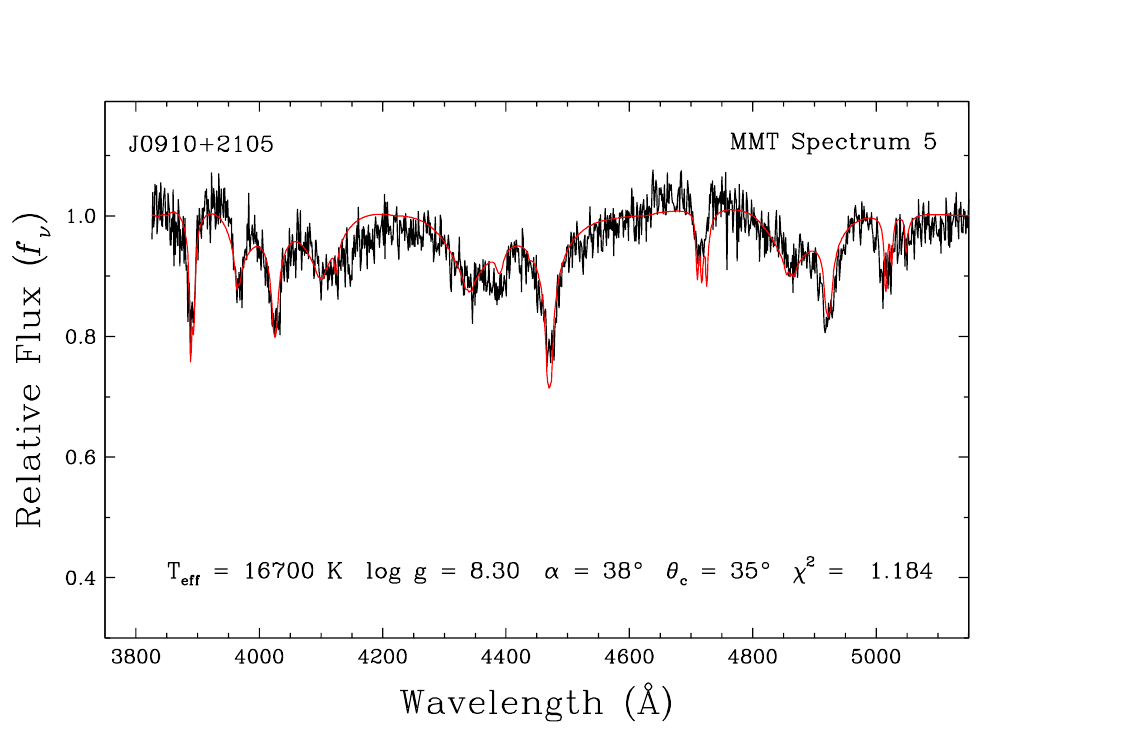}
    \caption{Top row: fits to both SDSS spectra using radiative atmospheres and assuming $\alpha = 30\degree$. Middle row: fits to the second and final spectra from APO on UT 2023 April 17th. Bottom row: fits to the fourth and final spectra from the MMT. Since $\alpha = 90\degree$ means pole-on, more prominent H lines are explained by a larger $\alpha$ angle.}
    \label{fig12}
\end{figure*}

Figure \ref{fig12} shows our model fits to several spectra with a fixed H cap size and varying viewing angle $\alpha$ between the plane of the sky and the magnetic axis. The top panels show the model fits to the two SDSS spectra, which are best explained by models with  $\alpha=53\degree$ and $38\degree$, respectively. Since $\alpha = 90\degree$ is for a pole-on configuration, more prominent H lines mean a larger $\alpha$ value. We see a broader range of H line depths in the APO spectra, and the example APO data shown in the middle panels are best explained with $\alpha$ values ranging from 51 to $17\degree$, whereas the MMT spectra shown in the bottom panels indicate $\alpha$ values of 54 to $38\degree$. Fitting all available spectra on the system, we find that $\alpha$ ranges from 13 to
$55\degree$ in this system. 

In the oblique rotator model, $\beta$ is defined as the angle between the magnetic and rotation axes, and $i$ is the angle between the rotation axis and the line of sight. Hence, as the star rotates, the angle between the magnetic axis and the line of sight ranges from $\beta - i$ to $\beta + i$ \citep[see Figure 2 of][]{Bailey11}. Since $\alpha$ is the angle between the plane of the sky and the magnetic axis, this means that $90-\alpha$, the angle between the line of sight and the magnetic axis, changes from $\beta - i$ to $\beta + i$ as the star rotates. Hence, the observed range of $\alpha=13-55\degree$ provides a unique solution for the system geometry, including unique values of $\beta$ and $i$. 

\begin{figure}
    \centering
    \includegraphics[width=3.6in,clip=true,trim=0.65in 0in 0.6in 0.6in]{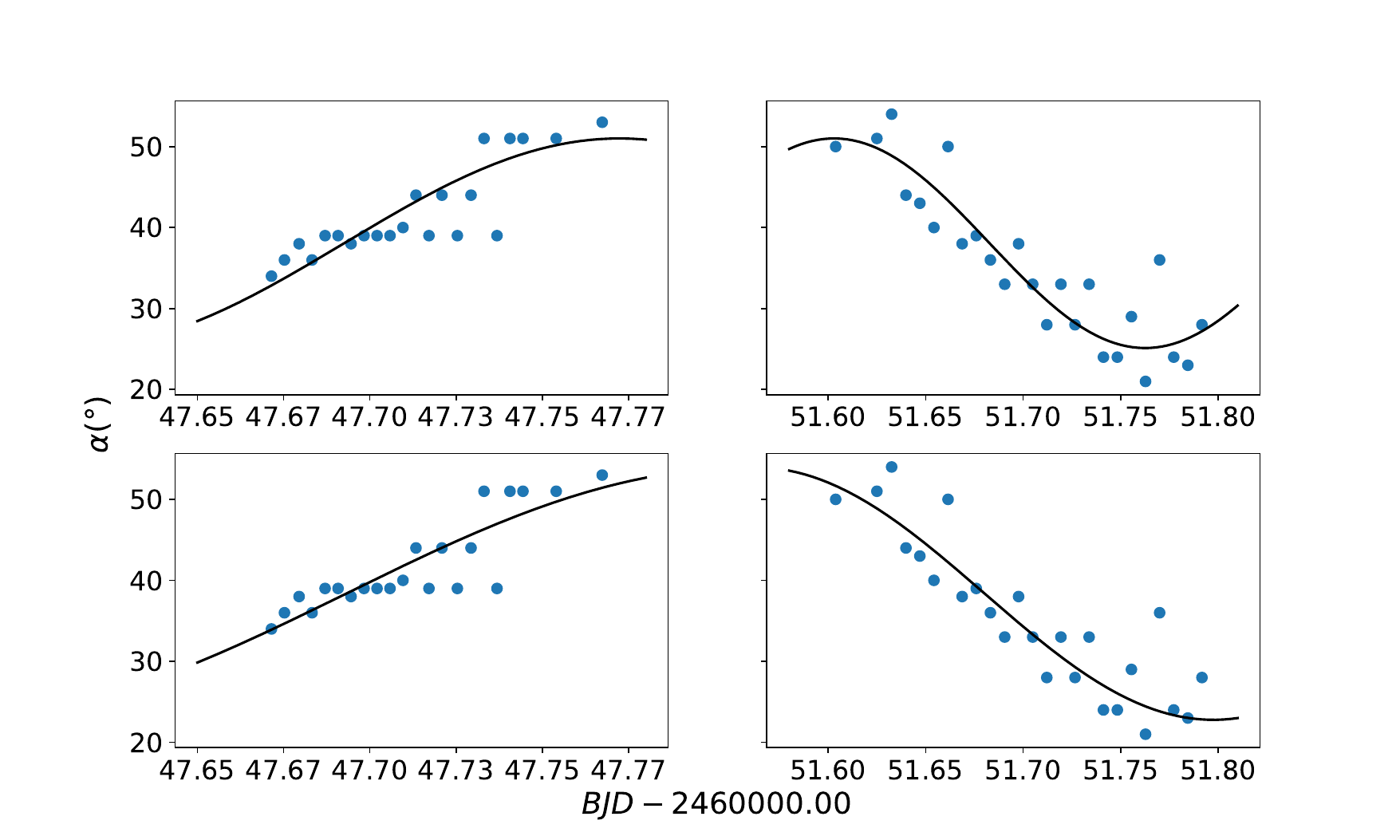}
    \caption{Top row: best-fit sine curve (black line) assuming a 7.7 hr rotation period with the $\alpha$ angle values (blue points) from the UT April 13th (left column) and UT April 17th (right column) APO data. Bottom row: same as the top row but with a 11.3 hr solution.}
    \label{fig13}
\end{figure}

The lower limit of $\alpha=13\degree$ comes from a low S/N spectrum from APO on UT May 26. Instead of relying on a single spectrum to mark its limits, we use all available spectra and fit a sine curve to the best-fitting $\alpha$ values. Figure \ref{fig13} shows the best-fitting $\alpha$ from our first two nights of APO data, which have the longest baseline out of all our data, and the sine curves corresponding to the 7.7 and 11.3 hour solutions. Both solutions match the data relatively well, so we are unable to remove the degeneracy in the rotation period.

For the 7.7 hr solution, the minimum and maximum correspond to $\alpha \approx 25\degree, 51\degree$ respectively. For the 11.3 hr solution, the values are $\alpha \approx 23\degree, 54\degree$. Combining these values with the definition of $\alpha$ above, we estimate both $\beta$ and $i$. The spectroscopic variations seen in J0910+2105 can be explained by the rotation of a star with the polar caps covering $\sim$18\% of the surface area (with $\theta_c=35\degree$), the angle between the rotation axis and the magnetic axis of $\beta=52\degree$, and the angle between the line of sight and the rotation axis of $i=13\degree$ for the 7.7 hr solution. For the 11.3 hr period, $\beta=52\degree$ and $i=16\degree$.

\section{Discussion}

\subsection{The Patchy Atmosphere}

We have successfully applied a magnetic patchy atmosphere model to our time-resolved spectra of J0910+2105. We show that the oblique rotator model with H polar caps provides an excellent match to the observed changes on the strength of the H lines in the spectra. However, there are clear degeneracies in the geometry of these fits, as the surface H distribution may be more complicated than just polar caps. Additionally, the fact that we see H and He at all phases means we require a more complicated distribution than something like Janus in \citet{Caiazzo23} where the lines completely vanish at specific phases, implying either side is dominated by one element. Regardless, while our fits may not return the exact values or geometry of what the atmosphere of J0910+2105 truly looks like, we have successfully produced a working model that yields excellent fits to the data. It is likely our current interpretation represents only a subsample of the various possible surface H/He abundance distributions. 

We can confidently rule out the possibility of J0910+2105 being in a DA+DB binary system. For that scenario to work, both stars would need to be magnetic which is quite rare based on population synthesis calculations from \citet{Briggs15}. We also do not see signs of Doppler shifts in our spectra, and a binary system would not produce the particular variations we see. While \citet{Genest19b} obtained a better fit using a DA+DB model than their homogeneous atmosphere model, the predicted He line profiles are consistently deeper compared to what is observed. With our patchy atmosphere model, we obtain significantly better fits to the depths of the line profiles. 

Most DBAs can be successfully modelled using a simpler model than the one we invoked here. A major question is what caused the inhomogeneities in the atmosphere of J0910+2105? The current paradigm to explain the re-appearence of DBs below $T_{\rm eff}\sim 30,000$ K is that DA white dwarfs with adequately thin surface H layers are converted to DBs through convective instability of the underlying He layer. The He layer mixes in with H above, turning the star into a DBA or DB white dwarf, depending on the total amount of H in the stellar envelope \citep{Bedard23}. 

The question then becomes why this mixing did not occur over the entire surface of the star. Because J0910+2105 is magnetic and because the motions of the material are confined along the magnetic field lines, this field can suppress convection and lead to varying elemental abundances throughout the atmosphere \citep[see for example,][]{Achilleos92}. Specifically, mixing is expected to be inefficient in regions where the field lines are tangential to the stellar surface (i.e. equatorial regions), and more efficient where the lines are radial to the surface (polar caps). \citet{Bedard23} successfully reproduced the range of atmospheric H abundances among cool DBAs using a deep H reservoir. With this reservoir, the magnetic field would carry H from the interior to the aforementioned regions where the field lines are radial to the surface. This process would result in H caps, matching the geometry we have invoked to model the atmosphere of J0910+2105. 

It is interesting to analyze this target in light of \citet{Hardy23}, who analyzed all known DBs with signs of Zeeman-splitting in the neutral He lines. In their original sample of 79 objects, only two were classified as DBAs, and were not analyzed as a result. It therefore seems quite rare to find a magnetic DBA despite DBAs themselves making up a large fraction of the DB population. J0910+2105 has given us an opportunity to observe the effects a magnetic field has on the distribution of elements in the mixed H/He atmosphere.   

\subsection{The Rotation Period and Magnetic Field}

While we have observed J0910+2105 long enough to detect visible variations in the strength of the H lines, the true rotation period is likely longer than our current baseline, and so we can only partially constrain the target's rotation period. Using the fitting process described in Section 3.1, we obtain a period of 7.7 or 11.3 hours. We expect that our APO observations on UT 2023 April 17th covered approximately half of one rotation given the significant weakening of H$\beta$ throughout the night. Hence a rotation period of roughly twice this baseline is highly likely. This period is also consistent with other magnetic white dwarfs. \citet{Kawka20} analyzed the distribution of rotation periods for known magnetic white dwarfs and found that the peak is located around 2 $-$ 3 hours. 

The question then becomes: what is the origin of this magnetic field, and how can the rotation period help us narrow it down? There are three potential origins to consider that can result in a magnetic white dwarf. The first is a fossil origin via flux conversion as the progenitor star evolves from a giant star to a white dwarf \citep{Tout04}. We use the initial-final mass relation from \cite{Cummings18} to estimate the progenitor mass of J0910+2105. With a white dwarf mass of 0.78 $M_\odot$, we obtain a 3.2 $M_\odot$ progenitor. \citet{Hermes17} analyzed the rotation rates of pulsating white dwarfs and found that those with masses from 0.51 $-$ 0.73 $M_\odot$ have an average rotation period of 35 hours, with a large standard deviation of 28 hours. Given J0910+2105's period, a fossil origin of the magnetic field cannot be ruled out. 

A second possible origin is via a double-degenerate merger or some form of binary evolution \citep{Liebert05,Liebert15}. Given an average mass of $\sim$0.8 $M_\odot$ for magnetic white dwarfs, this is a plausible channel to explain the magnetic white dwarfs with strong magnetic fields \citep{Ferrario15,Briggs15,Briggs18}. Additionally, a merger product or an isolated white dwarf that had undergone some binary evolution is likely to be spun up \citep{schwab21}. \citet{Moss23} analyzed seven magnetic white dwarfs and found that three have rotation periods less than an hour, one has a period of only a couple hours, and two are too low mass to have formed via single star evolution. J0910+2105 has an average mass for a magnetic white dwarf and an average period. \citet{Hardy23} also found that their magnetic DBs are more massive on average than typical DBs \citep{Genest19a}, so a binary formation channel is also possible. 

Finally, core crystallization can trigger a dynamo effect that can then induce a magnetic field \citep{Isern17,Schreiber21}. This does not apply to J0910+2105 though. Based on Figure 17 from \citet{Caron23}, who analyzed the role of crystallization in generating magnetic fields in cool white dwarfs, it is clear that J0910+2105 is much too hot to have begun crystallization for its mass.

\citet{Bagnulo22} identified two distinct populations of magnetic white dwarfs based on mass. The most massive white dwarfs, with magnetic field strengths on the order of 100 MG, are likely merger products given their high mass and field strength, and tendency to have very short rotation periods. On the other hand, detectable magnetic fields among average mass white dwarfs (0.5 to $0.7~M_{\odot}$) are very rare for the first Gyr of cooling, and the fields detected are generally weak, typically tens of kG. The frequency and field strength of these lower mass magnetic white dwarfs actually increases with cooling age. Hence, J0910+2105 is somewhat of an anomaly, being a very young but clearly magnetic white dwarf with a relatively large field, in spite of having a mass not far from the normal white dwarf mass.

\subsection{Comparison with other Patchy Atmosphere White Dwarfs}

Several other white dwarfs are known to show spectroscopic variations due to patchy atmospheres. The first is GD 323, a DAB found to have periodic variations in its line profiles by \citet{Pereira05}. Unlike J0910+2105, the He line strength varies out of phase with the H lines over $\sim$3.5 hours. \citet{Koester94} were unable to fit their spectra with homogeneous models, however they did not detect variations in their data. Since GD 323 lies just below the DB gap, \citet{Koester94} did consider a simple spot model to test the idea of an inhomogeneous atmosphere. \citet{Beauchamp93} used a more sophisticated model similar to ours, with an equatorial belt and polar caps, and obtained better results overall than the spot model. There are no signs of Zeeman-splitting in the spectra of GD 323 however, suggesting that any inhomogeneities are due to a very weak field, or more likely to the convective dilution process itself.

In contrast, Feige 7 is a DBA with a peculiar atmospheric structure and a strong magnetic field. \citet{Achilleos92} modelled this white dwarf with a pure He cap, followed by a pure H ring from latitudes $\sim$$100\degree$ and $\sim$$130\degree$ from the He pole, and then a He-dominant region throughout the rest of the atmosphere. The authors initially propose an accretion-based origin for this H ring, but rule it out given the short coupling distance (for the ionized accreting material to couple onto the magnetic field lines) in this region of the white dwarf. Instead, Feige 7 likely began convective dilution as it cooled and exited the DB gap. But due to the magnetic field, the dilution process did not distribute the He across the entire surface as in most DBs. \citet{Achilleos92} used an offset dipole model to derive a field strength $B_d = 35$ MG, which would clearly suppress convection, and an offset $a_z = 0.15$. The offset nature of the dipole in Feige 7 could further lead to complexities in the atmosphere given that the field strength across the surface is likely to vary significantly. While Feige 7 has a much stronger magnetic field than J0910+2105, the field strength we obtain is still strong enough to inhibit convection and prevent an even distribution of He across the surface. 

\citet{Caiazzo23} discovered both photometric and spectroscopic variations in a rapidly-rotating mixed atmosphere white dwarf, ZTF J203349.8+322901.1 (Janus). Here the H and He lines completely vanish and reappear as Janus rotates with a $\sim$15 minute period, suggesting one side is H-dominant and another is He-dominant. Similar to GD 323, there is no evidence of Zeeman-splitting in any of the absorption lines. The authors suggest two possible models to explain the stark differences in the atmosphere. The first is the aforementioned dilution model with a magnetic field strong enough to suppress convection but weak enough to be undetected. Their derived temperature of 34,900 K for the H side and 36,700 K for the He side suggests Janus is exiting the DB gap on the way to becoming a DB and undergoing this dilution process. The second involves the diffusion of H towards regions of high magnetic pressure and low gas pressure. If the magnetic field is stronger on one side, the magnetic pressure would be higher at the pole, and the ion pressure gradient would cause H to diffuse towards the pole. Both models however require a magnetic field that is at least a few tens of kG, which is observationally difficult to detect. Additionally, \citet{Caiazzo23} derived a large mass of 1.27 $M_\odot$ assuming a carbon-oxygen core and 1.21 $M_\odot$ for an oxygen-neon core. Given the large mass and short spin period, Janus is thus likely a merger product, which could further complicate the process that creates the double-faced atmosphere.

\citet{Wesemael01} determined that LB 8915 (erroneously called LB 8827), previously known to be a hot DB \citep{Wesemael95}, is actually a weakly magnetic, variable DBA. Their derived $T_{\rm eff}$ range of 21,300 $-$ 27,700 K places this object within or near the instability strip for pulsating DBs. They did not detect photometric variations, but they do detect changes in the H line strengths. While \citet{Wesemael01} detect a magnetic field in LB 8915, it is both weak and unconstrained; they made three spectropolarimetric observations of LB 8915, with resulting fields of $B_e = 67 \pm 40, 108 \pm 24$, and $24 \pm 13$ kG. It is possible the dipole is offset, creating variations in the field strength, however more observations are needed to confirm this. 

Lastly, \citet{Pshirkov20} detected photometric variations in the DBA WD 1832+089 due to rapid rotation. The derived mass of $1.33\pm0.01$ $M_\odot$ and period of 353 s strongly suggests a past merger occurred to form this white dwarf. Since the variations are due to rotation, it is likely this target also has an inhomogeneous atmosphere, though again no Zeeman splitting is seen in the spectra.

In light of this discussion, it would seem J0910+2105 is most similar to Feige 7 out of the targets we have listed. It is the only one that shows clear Zeeman-splitting alongside J0910+2105. \citet{Hardy23} derived an effective temperature of $T_{\rm eff} = 20848 \pm 1077$ K and a mass of $1.13 \pm 0.18$ $M_\odot$ for Feige 7, although they did not consider H in their models. This puts J0910+2105 at a slightly lower temperature and mass. 

Further analysis of magnetic DB or DBA candidates could aid in finding common threads between these objects by expanding this currently small sample. Following up on the the remaining nine unresolved DA+DB candidates from \citet{Genest19b} is a natural starting point. Even if Zeeman-splitting is not detectable in the spectra, it is possible these objects have inhomogeneous atmospheres that would manifest as spectral variations with time-series spectroscopy.

In summary, we have detected variations in the strength of the H lines in the DBA white dwarf SDSS J091016.43+210554.2. We obtained time-series spectroscopy at the APO 3.5m telescope and the 6.5m MMT and estimate the rotation period to be either 7.7 or 11.3 hours. We use a magnetic patchy atmosphere model to fit our time-series spectroscopy. We obtain excellent fits using the oblique rotator model, and determine the angle between the magnetic and rotation axis ($\beta = 52\degree$) and the angle between the rotation axis and line of sight ($i = 13$ - $15.5\degree$). The derived $T_{\rm eff}$ of 16,747 K places this target at a cooler temperature outside of the DB gap. The magnetic field ($B\approx0.5$ MG) appears to be strong enough to affect convective energy transport in a way that H is brought to the surface from the deep interior more efficiently along the magnetic field lines, thus creating a patchy atmosphere with H polar caps. With the use of our polar cap model, we have successfully shown how the magnetic field affects the distribution of elements in the white dwarfs atmosphere. 

\section{Acknowledgements}
We thank the referee, John Landstreet, for feedback that significantly improved this article.
We are grateful to Antoine Bédard for useful discussions. We also thank Stefan Jordan for providing the magnetic He line models.
We thank the students of the Spring 2023 Advanced Observatory Methods class at OU, Cosme Aquino Ovelar, Michael Bartlett, Raelin Lane, Rachel Smith, and Sean Smith for obtaining spectra of J0910+2105 during their telescope training on UT 2023 April 13.
This work is supported in part by the NSF under grant AST-2205736, the NASA under grant 80NSSC22K0479, the NSERC Canada, the Fund FRQ-NT (Qu\'ebec), and by the Smithsonian Institution. 

Based on observations obtained with the Apache Point Observatory 3.5-meter telescope, which is owned and operated by the Astrophysical Research Consortium. Observations reported here were obtained at the MMT Observatory, a joint facility of the Smithsonian Institution and the University of Arizona. 

This paper includes data collected with the TESS mission, obtained from the MAST data archive at the Space Telescope Science Institute (STScI). Funding for the TESS mission is provided by the NASA Explorer Program. STScI is operated by the Association of Universities for Research in Astronomy, Inc., under NASA contract NAS 5–26555.

\section*{Data availability}
The data underlying this article will be shared on reasonable request to the corresponding author.

\newcommand{\newblock}{}
\bibliographystyle{mnras}
\bibliography{MossDBA.bib}

\end{document}